%% file: main.tex
\pdfoutput=1
\documentclass[11pt]{article}

\usepackage[final]{acl}

\usepackage{times}
\usepackage{latexsym}

\usepackage[T1]{fontenc}

\usepackage[utf8]{inputenc}

\usepackage{microtype}

\usepackage{inconsolata}

\usepackage{graphicx}

\usepackage{amsmath}
\usepackage{amssymb}
\usepackage{mathtools}
\usepackage{amsthm}
\usepackage[most]{tcolorbox}

\usepackage[capitalize,noabbrev]{cleveref}

\theoremstyle{plain}

\theoremstyle{definition}

\theoremstyle{remark}

\newcommand{\cready}[1]{#1}

\usepackage{xcolor}
\usepackage{microtype}
\usepackage{graphicx}
\usepackage{subcaption}
\usepackage{booktabs} %
\usepackage{makecell}
\usepackage[most]{tcolorbox}
\usepackage{pgfplots}
\usepgfplotslibrary{polar}
\pgfplotsset{compat=1.18}

\definecolor{tabblue}{RGB}{31,119,180}
\definecolor{taborange}{RGB}{255,127,14}
\definecolor{tabgreen}{RGB}{44,160,44}
\definecolor{tabred}{RGB}{214,39,40}

\DeclareUnicodeCharacter{2212}{-}

\newtcolorbox{promptbox}[2][]{%
    enhanced,
    unbreakable,
    colback=gray!5!white,
    colframe=gray!60!black,
    fonttitle=\bfseries,
    title={#2},
    attach boxed title to top left={yshift=-2mm, xshift=3mm},
    boxed title style={colback=gray!60!black},
    fontupper=\small\ttfamily,
    #1
}

\title{Long Listening Thoughts: Eliciting Open Auditory Reasoning with \\ Deliberative Perception and Cognitive Refinement}

\author{
  \quad\textbf{Jaeyeon Kim\textsuperscript{1}}
  \quad\textbf{Chao-Han Huck Yang\textsuperscript{2}}
  \quad\textbf{Luoyi Zhang\textsuperscript{1}}
  \quad\textbf{Chan-Jan Hsu\textsuperscript{1}} \\
  \quad\textbf{Fernando Ruiloba Portilla\textsuperscript{1}}
  \quad\textbf{Jinchuan Tian\textsuperscript{1}}
  \quad\textbf{Shinji Watanabe\textsuperscript{1}}
  \quad\textbf{Carlos Busso\textsuperscript{1}} \\
  \textsuperscript{1}Carnegie Mellon University
  \quad \textsuperscript{2}NVIDIA \\
  {\small \texttt{jaeyeon2@andrew.cmu.edu
  \quad huckiyang@ieee.org \quad luoyiz@andrew.cmu.edu}} \\
  {\small \texttt{chanjanh@andrew.cmu.edu \quad feruiloba@gmail.com \quad jinchuat@andrew.cmu.edu}} \\
  {\small \texttt{shinjiw@ieee.org \quad busso@cmu.edu}}
}

\begin{document}
\maketitle
\begin{abstract}
We introduce Long Listening Thoughts (LLT), a framework for constructing post-training datasets that improves audio reasoning by pairing deliberate perceptual grounding with explicit cognitive behaviors.
On audio understanding and reasoning benchmarks, large audio–language models (LALMs) are limited primarily by perceptual errors, yet existing post-training datasets rely on audio descriptions of limited diversity.
Through a four-stage pipeline, LLT (1) builds multi-aspect dense audio descriptions spanning semantic, acoustic, and temporal information, (2) generates more challenging questions, (3) synthesizes incorrect reasoning traces, and (4) extends short reasoning traces into long ones via a thought-continuation strategy that elicits cognitive behaviors such as verification and backtracking.
Using LLT, we synthesize 1.6M reasoning traces and train Qwen2.5-Omni-7B with a two-stage pipeline of supervised fine-tuning (SFT) followed by direct preference optimization (DPO).
This improves accuracy on MMAU Test-mini ($71.5 \rightarrow 78.2$), MMAR ($56.7 \rightarrow 63.7$), and MMSU ($60.6 \rightarrow 66.0$).
Project page: \href{https://jaeyeonkim99.github.io/llt}{https://jaeyeonkim99.github.io/llt}
\end{abstract}

\input{1_introduction/introduction}
\input{2_related_work/related_work}
\input{3_method/method}
\input{5_experiment/experiment}

\section{Conclusions}
We introduce Long Listening Thoughts (LLT), a framework for constructing post-training datasets for audio reasoning that explicitly targets perceptual grounding and cognitive refinement. By integrating multi-aspect dense audio descriptions with a thought-continuation strategy, LLT produces 1.6M reasoning traces that encourage reflective and corrective reasoning behaviors. Extensive experiments show that models trained with LLT achieve improved audio reasoning across multiple benchmarks. \cready{We are sharing the dataset, models, and full data construction pipeline to support reproducibility and facilitate future research}.

\section*{Limitations}

While LLT consistently improves audio reasoning, several limitations should be acknowledged.
First, LLT enriches an existing audio CoT dataset rather than collecting new audio, and our study focuses on clips of up to 30 seconds. This design keeps the pipeline scalable and compatible with prior datasets, but the resulting gains may not transfer directly to substantially longer recordings or to domains underrepresented in the source data. Extending the pipeline along these axes is a natural direction for future work.

Second, following prior audio reasoning data-construction pipelines~\cite{audsemthinker, audio_mcq}, LLT relies largely on model-based generation and verification. We complement this with human inspection on a sampled subset (Appendix~\ref{app:human_eval}), but such automated verification cannot fully guarantee the correctness and quality of every trace at scale.

Finally, the fine-grained perceptual capabilities that LLT strengthens, such as acoustic attribute recognition, temporal localization, and speaker diarization, are intended to support reasoning, but in principle could also be applied to infer sensitive information from audio. To mitigate this, we build only on publicly available research datasets.

\section*{Acknowledgments}
\cready{This material is based upon work supported in whole or in part with funding from the Department of Defense. Any opinions, findings, conclusions, or recommendations expressed in this material are those of the author(s) and do not necessarily reflect the views of the DoD and/or any agency or entity of the United States Government.}

\bibliography{main}

\appendix
\input{7_appendix/appendix}

\end{document}

%% file: 1_introduction/introduction.tex
\section{Introduction}
\label{sec:intro}

Recent progress in large audio–language models (LALMs) has substantially advanced audio processing across a wide range of domains, including speech, sound events, music, or mixed acoustic scenes \cite{deshmukh2023pengi,radhakrishnan2023whispering, gong2024ltu, ghosh2024gama, qwen2_audio, af3, abouelenin2025phi, tian2025opuslm, bagpiper}. These models support diverse objectives ranging from traditional tasks such as automatic speech recognition (ASR) to tasks that require complex reasoning such as audio question answering \cite{yang2025multi, interspeech_reasoning_challenge}. To further enhance their reasoning capabilities, recent efforts have focused on post-training strategies such as constructing chain-of-thought (CoT) datasets \cite{audio_reasoner, af_cot, audsemthinker} for supervised fine-tuning (SFT), applying reinforcement learning (RL) techniques \cite{r1_aqa, omni_r1, audio_thinker, cesar}, or combining both approaches by applying RL after SFT \cite{sari, audio_mcq}.

\input{3_method/figure_approach}

However, despite these advances, LALMs still exhibit notable limitations when evaluated on audio understanding and reasoning benchmarks. Recent benchmarks covering a broad range of audio reasoning tasks \cite{mmar, mmau, mmsu} show that perceptual errors account for the majority of failures, followed by reasoning errors. Moreover, LALMs are known to struggle with perceiving fine-grained details in audio, such as low-level acoustic attributes \cite{wow_bench} and precise temporal structure \cite{timeaudio}.
Existing post-training approaches largely rely on audio descriptions generated by the LALMs themselves~\cite{chenaudio, hu2024large} or on ground-truth metadata \cite{lu2025desta2, audio_mcq} with limited diversity (e.g., coarse audio event labels in AudioSet \cite{audioset}) when constructing training data. This reliance constrains the model’s ability to reason over richer and more deliberate perceptual cues that go beyond its original perceptual capacity. These observations suggest that post-training datasets guided by more deliberate, multi-aspect perceptual information could substantially improve audio reasoning performance.

In parallel, test-time scaling has been shown to be effective for complex reasoning tasks such as math and coding \cite{deepseek,jaech2024openai}. However, explicit strategies for leveraging longer reasoning traces in audio reasoning remain relatively underexplored.
Prior work indicates that successful test-time scaling is closely related to structured cognitive behaviors such as verification and backtracking~\cite{cognitive, yeo2025demystifying} that enable models to explore alternative hypotheses and refine their reasoning. These behaviors have been further shown to be effective in perceptually oriented visual reasoning \cite{lpt, lgt}, yet have not been systematically incorporated into audio reasoning frameworks.

Motivated by these observations, we propose Long Listening Thoughts (LLT), a new framework for constructing post-training datasets for audio reasoning that integrates deliberate multi-aspect perceptual grounding with explicit cognitive behaviors. Building upon existing audio CoT datasets, LLT introduces a scalable four-stage pipeline in which each stage targets a distinct factor underlying high-quality audio reasoning: \textbf{Stage~1} constructs multi-aspect dense audio descriptions to strengthen perceptual grounding, \textbf{Stage~2} augments more challenging questions that demand non-trivial reasoning, \textbf{Stage~3} synthesizes plausible incorrect reasoning traces that provide a corrective signal for reflection and correction, and \textbf{Stage~4} generates long reasoning traces with explicit cognitive behaviors. The overview is illustrated in Figure~\ref{fig:approach}.

Using the LLT framework, we construct a post-training dataset comprising approximately 1.6M reasoning traces, supporting a two-stage training pipeline of SFT followed by Direct Preference Optimization (DPO) \cite{rafailov2023direct}. \cready{We are sharing the dataset, trained models, and the full data construction pipeline to support reproducibility and future research.}
Through extensive analysis, we demonstrate the effectiveness of the proposed four-stage pipeline in generating perceptually grounded and cognitively structured reasoning traces.
When training Qwen2.5-Omni-7B \cite{qwen25_omni} with LLT, we observe substantial improvements over the original model on audio reasoning benchmarks.

%% file: 3_method/figure_approach.tex
\begin{figure*}[ht]
    \centering
    \includegraphics[width=\textwidth]{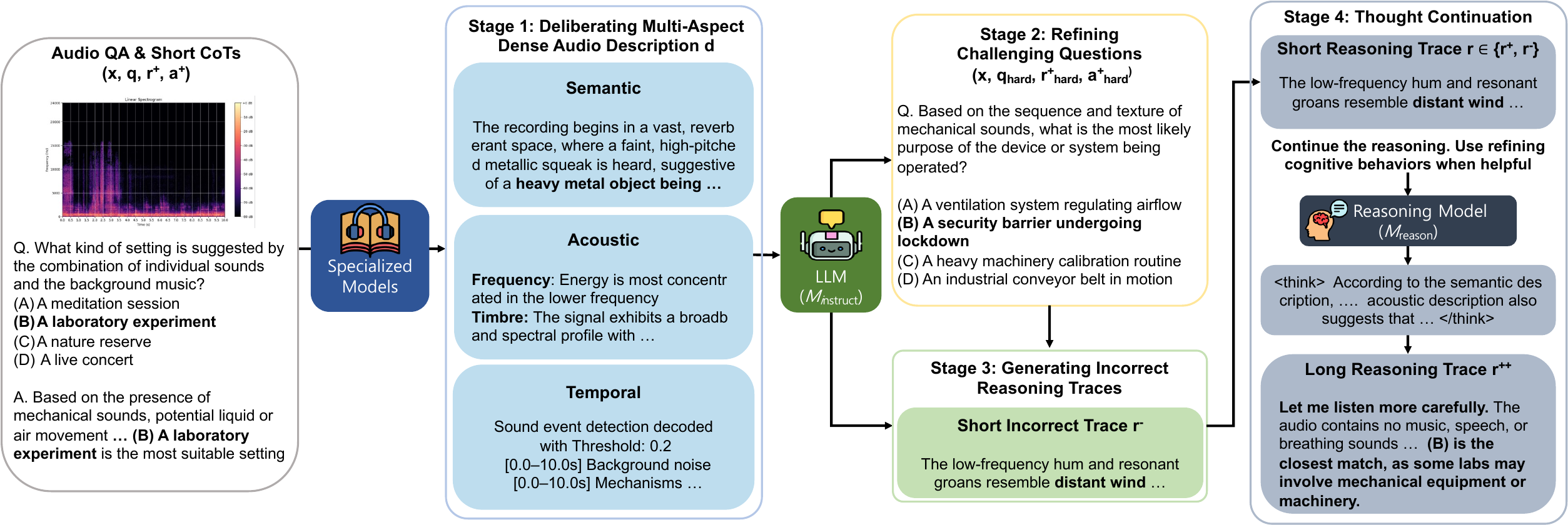}
    \vspace{-\baselineskip}
    \caption{\textbf{Overview of the proposed \emph{Long Listening Thoughts} pipeline.} \textbf{Stage 1} extracts multi-aspect audio descriptions
(semantic, acoustic, temporal) from specialized models. \textbf{Stage 2} augments more challenging questions using an LLM. \textbf{Stage 3} generates incorrect reasoning traces.
\textbf{Stage 4} extends traces via thought continuation to elicit cognitive behaviors.}

\label{fig:longlisteningthought_overview}

    \label{fig:approach}
\end{figure*}

%% file: 2_related_work/related_work.tex
\section{Related Work}
\textbf{Large Audio Language Models.} Recent advances in LALMs have demonstrated impressive capabilities in processing diverse sound domains, including speech, sound events, and music \cite{deshmukh2023pengi, qwen2_audio, kimi_audio, ghosh2024gama, af3, gong2024ltu, lu2025desta2, tian2025opuslm, bagpiper}. Moreover, recent omni-modal models process audio inputs alongside visual inputs by integrating encoders from multiple modalities with an LLM. These approaches have also demonstrated strong performance in audio understanding \cite{qwen25_omni, qwen3_omni, omni_vinci}.

To systematically evaluate LALMs' capabilities, several benchmarks assess general audio understanding and reasoning across a wide range of tasks~\cite{mmau, mmar, mmsu, mmau_pro}. Analyses on these benchmarks show that audio reasoning remains a challenging task for LALMs, and that perceptual errors account for the majority of failures, followed by reasoning errors~\cite{mmau, mmar}. In this work, we propose a post-training framework that addresses this limitation by constructing reasoning traces grounded in richer, multi-aspect perceptual information and incorporating cognitive behaviors such as verification and backtracking, improving both perception and reasoning.

\noindent\textbf{Audio Reasoning Models.} To further enhance the reasoning capabilities of LALMs, recent works have proposed a variety of approaches targeting audio reasoning more explicitly. One direction leverages chain-of-thought reasoning traces, either through prompting~\cite{audio_cot} or training. On the training side, prior work either folds CoT data into pre-training~\cite{af3} or curates dedicated CoT datasets for SFT~\cite{audio_reasoner, af_cot, audsemthinker}. Across these efforts, however, the audio supervision is typically drawn from high-level event captions or metadata, providing only limited perceptual grounding. LLT departs from this by grounding traces in multi-aspect perceptual information, while extending prior CoT datasets through a scalable pipeline rather than constructing supervision from scratch.

Another direction applies online RL, typically Group-Relative Policy Optimization (GRPO)~\cite{shao2024deepseekmath}, on top of audio question-answering data. Several works apply RL directly to LALMs~\cite{omni_r1, r1_aqa, ke-omni-r, audio_thinker}, while others prime reasoning with SFT before training with GRPO~\cite{sari, audio_mcq}. They aim to improve training curricula~\cite{sari, audio_mcq} or reward design~\cite{cesar, audio_thinker} for GRPO. Rather than targeting the GRPO stage, LLT focuses on the upstream SFT stage to install perceptual grounding and cognitive behaviors. Specifically, we propose a post-training dataset and an SFT $\rightarrow$ DPO training pipeline for this purpose, while the resulting model remains compatible with GRPO for further capability gains.

%% file: 3_method/method.tex
\section{Long Listening Thoughts}
\label{sec:method}
We propose Long Listening Thoughts (LLT), a framework for constructing post-training datasets of perceptually grounded and cognitively structured reasoning traces for audio reasoning. Rather than generating questions and reasoning traces from scratch~\cite{audio_reasoner, af3, audio_mcq}, LLT enriches examples drawn from an existing audio CoT dataset, which improves scalability and keeps the resulting data compatible with prior datasets.

\paragraph{Notation.}
We write an example of the source audio CoT dataset as $(x, q, r^{+}, a^{+})$, where $x$ is the audio, $q$ the question, $r^{+}$ the correct short reasoning trace, and $a^{+}$ the correct answer. Stage~1 attaches a dense audio description $d$, extending the tuple to $(x, q, r^{+}, a^{+}, d)$; when a component is irrelevant to a given step we write ``$-$'' in its place (e.g., $(-, q, r^{+}, a^{+}, d)$ for a text-only step that does not access the audio $x$). Superscripts mark correctness: $r^{-}$ and $a^{-}$ denote an \emph{incorrect} short trace and answer, respectively (Stage~3). A \emph{long} reasoning trace $r^{++}$ extends a short trace with explicit cognitive behaviors that revisit and refine the initial reasoning (Stage~4; e.g., \textit{``Wait, the initial reasoning incorrectly assumes\ldots''}). Two language models drive the construction pipeline: an instruction-following LLM $M_{\text{instruct}}$ and a reasoning LLM $M_{\text{reason}}$.

\paragraph{Overview.}
Given a source example, LLT applies the four stages illustrated in Figure~\ref{fig:approach}: \textbf{Stage~1} builds the multi-aspect dense audio description $d$ (Section~\ref{sec:stage1}); \textbf{Stage~2} rewrites the question into a more challenging one (Section~\ref{sec:stage2}); \textbf{Stage~3} synthesizes an incorrect short trace $r^{-}$ (Section~\ref{sec:stage3}); and \textbf{Stage~4} continues a short trace $r \in \{r^{+}, r^{-}\}$ into a long trace $r^{++}$ through thought continuation (Section~\ref{sec:stage4}). These stages yield three trace types per example: short correct $r^{+}$, \emph{correct-to-refine} $(r^{+}\!\rightarrow\! r^{++})$, and \emph{incorrect-to-correct} $(r^{-}\!\rightarrow\! r^{++})$, where the arrow denotes thought continuation; their roles in training are described in Section~\ref{sec:multi-stage}.

\subsection{Stage 1: Deliberating Multi-Aspect Dense Audio Description}

\label{sec:stage1}

As discussed in Section~\ref{sec:intro}, perceptual errors are the dominant failure mode on audio reasoning benchmarks, motivating richer perceptual grounding. However, existing CoT datasets typically rely on audio metadata \cite{audio_mcq, lu2025desta2} or captions generated by LALMs themselves, such as Qwen2-Audio \citep{qwen2_audio, sari, audio_thinker}, which capture only limited aspects of the audio. One notable exception, AudSemThinker \cite{audsemthinker}, aggregates audio and visual analyses from multiple models, but still relies on Qwen2-Audio, audio captioning \cite{cone_tte}, and event tagging \cite{beats} models that convey overlapping, high-level semantic information.
Therefore, we propose to build a multi-aspect dense description for each audio input, providing a more holistic view of the auditory scene and richer cues for reasoning trace generation.

Figure~\ref{fig:approach} highlights the three complementary aspects of the audio information extracted in Stage~1: (i) semantic information describing high-level audio events and their interpretations, (ii) acoustic information capturing attributes below the event level, such as duration, pitch, timbre, and amplitude, and (iii) temporal information providing precise timestamps and temporal structure. For semantic information, we use Qwen3-Omni-Captioner \cite{qwen3_omni}, a recent specialized model to generate dense descriptions of audio events. For acoustic information, we directly leverage rich acoustic cues from spectrograms by prompting a vision-language model (VLM) to describe attributes such as frequency distribution, duration, spectral patterns, and reverberation. We find that providing explicit grid and axis information that the VLM can numerically ground, together with detailed interpretation guidelines (e.g., which acoustic attributes to attend to and illustrative examples), substantially improves the quality of spectrogram-based descriptions. Since LALMs typically lack a detailed understanding of low-level acoustic attributes, this spectrogram-driven approach yields more accurate and informative acoustic descriptions than relying solely on audio–language models (see Appendix~\ref{appen:acoustic_example} for examples). We use Qwen3-VL-32B \cite{qwen3_vl} as the VLM for spectrogram interpretation.

For temporal information, we employ specialized models to obtain accurate timestamps and temporal structure.
We first select sub-tasks that collectively cover complementary aspects of temporal structure (speech, speaker turns, sound events, musical structure, and chord progression), and then choose strong-performing, open-source models for each.
Specifically, Whisper-large-v3 \cite{whisper} is used to generate timestamps for speech segments, while pyannote \cite{pyannote} provides speaker diarization timestamps. In addition, we use a pretrained sound event detection model \cite{pretrained_sed} based on BEATs \citep{beats} to extract event boundaries. We apply multiple detection thresholds (0.2 and 0.5) for sound event detection in order to capture both prominent and subtle events.
Finally, for audio clips detected to contain music, we additionally apply allin1 \cite{allin1} music structure analysis and autochord chord recognition \cite{autochord} to obtain temporally aligned musical annotations. By concatenating all information, we form the dense audio description $d$. An example of the generated dense description is shown in Appendix~\ref{appen:description_example}.

\subsection{Stage 2: Refining Challenging Questions}

\label{sec:stage2}

With a more holistic view of the auditory scene, we can generate more challenging questions grounded in richer perceptual evidence. Given a source datapoint $(x, q, r^{+}, a^{+})$, we prompt an instruction-following large language model $M_{\text{instruct}}$ to generate a harder datapoint $(x, q_{\text{hard}}, r^{+}_{\text{hard}}, a^{+}_{\text{hard}})$. This process is demonstrated in Stage 2 of Figure~\ref{fig:approach}. Specifically, the model is instructed to retain the original question format and required skill set, while increasing difficulty by making the question more audio-aware or by requiring more complex reasoning. For instance, a content-recall question can be rewritten to demand closer listening (\emph{``Summarize the speaker's advice''} $\rightarrow$ \emph{``How does the speaker's intonation distinguish hopeless failure from greater success?''}) or deeper inference (\emph{``What does the speaker claim about drama critics?''} $\rightarrow$ \emph{``What does the other speaker's sarcastic reply reveal about her attitude?''}). Additionally, we verify the generated questions using $M_{\text{instruct}}$ given the dense audio description $d$ to ensure that (i) the question can be solved by listening to the audio, (ii) the final answer is correct, and (iii) the associated reasoning trace is logically valid.
Empirically, when evaluated on 10k samples from AF-Think \cite{af3}, Qwen2.5-Omni-7B's accuracy drops from 72.5\% on the original questions to 63.0\% on their hardened counterparts, confirming that Stage 2 produces meaningfully harder questions.

\subsection{Stage 3: Generating Incorrect Reasoning Traces}

\label{sec:stage3}

To enhance reflective and corrective reasoning, we generate incorrect short traces $r^{-}$ that contrast with the correct $r^{+}$.
A natural source would be to directly sample from the LALM being trained by prompting it on questions and collecting its incorrect outputs, an approach explored in vision reasoning to build negative reasoning traces~\cite{lgt, lpt}.
However, this strategy has several limitations. First, it does not consistently generate incorrect reasoning traces even when explicitly prompted to reason~\cite{audio_thinker}. Second, the resulting errors lack sufficient diversity to reflect the range of misperceptions that can arise in real-world scenarios.
Third, traces sampled from a single model tie the negative supervision to that model's failure distribution, limiting the generalizability of the resulting dataset to other LALMs.

To address these limitations, we synthesize incorrect reasoning traces by prompting a language model with $(-, q, r^{+}, a^{+}, d)$.
Specifically, we instruct the model to retain the length and structural format of the correct reasoning trace $r^+$, while generating a plausible but incorrect reasoning trace $r^-$ that arises from misperceiving or misinterpreting fine-grained details in the dense audio description.
This process encourages the generation of errors that are subtle, realistic, and grounded in perceptual ambiguity.
This process is illustrated in Stage~3 of Figure~\ref{fig:approach}.
We generate $r^-$ for both original questions and the challenging questions produced in Stage~2.
The resulting incorrect reasoning traces are subsequently used in Stage~4 to generate long reasoning traces with explicit cognitive behaviors.
An ablation comparing different sources of incorrect reasoning traces is provided in Table~\ref{tab:ablation}.

\subsection{Stage 4: Thought Continuation}
\label{sec:stage4}
We augment short reasoning traces by appending continuations that exhibit explicit cognitive behaviors, building on the initial reasoning rather than replacing it. Specifically, we apply thought continuation to both correct traces $r^+$ from the source CoT dataset and Stage~2, and incorrect traces $r^-$ from Stage~3, yielding two types of long reasoning traces: \textit{correct-to-refine} traces $(r^+ \rightarrow r^{++})$ and \textit{incorrect-to-correct} traces $(r^- \rightarrow r^{++})$.

Previous work in the vision domain adopts a \textit{thought expansion} approach, distilling long reasoning traces from a large reasoning model $M_{\text{reason}}$ by combining a short trace $r$ with a cognitive trigger such as \textit{``Wait''} or \textit{``Hmm''} and prompting $M_{\text{reason}}$ to complete the trace within the \texttt{<think></think>} block~\cite{lpt, lgt}. Applying this directly to audio is ineffective. Due to the length and multi-aspect nature of dense audio descriptions, $M_{\text{reason}}$ struggles to determine which parts to attend to and produces excessively long, unfocused, or repetitive traces. This is undesirable for training LALMs, which should ground reasoning in the audio itself.

We instead adopt a \textit{thought-continuation} approach. Given an initial trace $r \in \{r^+, r^-\}$, a cognitive trigger $c$, and the dense audio description $d$ from Section~\ref{sec:stage1}, we provide $r \oplus c$ (where $\oplus$ denotes concatenation) in the prompt and have $M_{\text{reason}}$ generate a continuation of the reasoning. The model deliberates on how to continue $r \oplus c$ inside the \texttt{<think></think>} block and produces the final continuation in the subsequent \texttt{<answer></answer>} block, naturally refining its own reasoning while following the original output format. The resulting long trace $r^{++}$ keeps the initial reasoning $r$ intact and appends a continuation that revisits and refines it, reaching the correct answer whether the prefix is $r^+$ (correct-to-refine) or $r^-$ (incorrect-to-correct). Correctness is not guaranteed by construction; we enforce it through the verification step described below, which discards traces whose final answer is incorrect.

To encourage explicit cognitive behaviors within the continuation, we prompt $M_{\text{reason}}$ to encode verification and backtracking, and use both general triggers (\textit{Wait}, \textit{Hmm}, \textit{Alternatively}, \textit{Let me check again}) and audio-aware triggers (\textit{Let me listen again}, \textit{Let me listen carefully}) as $c$ to encourage grounding in the audio. This approach produces traces that are more concise and SFT-friendly than directly distilled thinking tokens; a comparison is shown in Appendix~\ref{appen:stage4_example}.

In addition, we apply verification during long reasoning trace generation. Specifically, we prompt $M_{\text{instruct}}$ with the generated continuation and $(-, q, a^{+}, d)$, and verify the following criteria: (i) whether the final answer is correct, (ii) whether the reasoning is logically valid, and (iii) whether the output format is valid (e.g., no references to the description text and proper use of the \texttt{<answer>} format). For criteria (i) and (ii), samples that fail verification are discarded. For criterion (iii), minor formatting errors are corrected with $M_{\text{instruct}}$ during verification, while samples with severe format violations are discarded. We complement this automated verification with human inspection on a sampled subset (Appendix~\ref{app:human_eval}).

\subsection{Two-Stage Training}
\label{sec:multi-stage}
The LLT pipeline produces three types of reasoning traces per question: $(x, q, r^{+}, a^{+})$, $(x, q, r^{+}\!\rightarrow\! r^{++}, a^{+})$, and $(x, q, r^{-}\!\rightarrow\! r^{++}, a^{+})$. While incorrect-to-correct traces $(r^{-}\!\rightarrow\! r^{++})$ from Stage~3 are essential for installing cognitive behaviors such as verification and backtracking, our analysis (Section~\ref{sec:analysis}) reveals that incorporating them directly into SFT hinders general reasoning performance, as the model struggles to distinguish intermediate $r^-$ from $r^+$ under standard SFT supervision. To address this, we disentangle the role of each trace type across two training stages. We use only $(x, q, r^{+}, a^{+})$ and $(x, q, r^{+}\!\rightarrow\! r^{++}, a^{+})$ for SFT to strengthen perceptual grounding and refined reasoning, and reserve $(x, q, r^{-}\!\rightarrow\! r^{++}, a^{+})$ for the subsequent DPO stage, where the preference structure naturally exploits the contrast between incorrect and corrected reasoning to induce cognitive behaviors. Unlike prior work \cite{lpt} that places $(r^{-}\!\rightarrow\! r^{++})$ directly in SFT, our two-stage design avoids this conflict and yields both stronger performance and more pronounced cognitive behaviors (Section~\ref{sec:analysis}).
Specifically, for each input $(x, q)$ we construct three types of preference pairs that prefer correct over incorrect reasoning: \emph{short-contrast} $(r^{+} \succ r^{-})$, \emph{refine-contrast} $(r^{+}\!\rightarrow\! r^{++} \succ r^{-})$, and \emph{correction-contrast} $(r^{-}\!\rightarrow\! r^{++} \succ r^{-})$. The correction-contrast pair specifically teaches the model to prefer trajectories that recognize and correct initial errors over those that commit to them.

\subsection{Implementation Details}
\label{sec:impl}
We use AF-Think as the initial short CoT dataset, as it covers diverse audio domains and scenarios. For reproducibility, we exclude licensed data such as Music4all~\cite{music4all} and limit audio duration to 30 seconds for compatibility with general LALM architectures, yielding 338k audio clips, each with one original question--answer pair. Following AF-Think, we retain both multiple-choice and free-form questions. Stage~2 augments these with 316k harder questions, and Stages~3--4 produce the three reasoning-trace types, giving 1.6M traces in total that we release to support diverse post-training setups. For our two-stage pipeline, we use the 1.16M short-correct and correct-to-refine traces for SFT and sample 60k DPO preference pairs (20k per pair type; Section~\ref{sec:multi-stage}). Detailed dataset statistics are reported in Table~\ref{tab:dataset_stats}.

We instantiate $M_{\text{instruct}}$ as Qwen3-Next-80B-Instruct and $M_{\text{reason}}$ as Qwen3-Next-80B-Thinking. All pipeline components, including the Stage-1 submodules, are open-source, ensuring full reproducibility. The prompts used at each stage are provided in Appendix~\ref{appen:prompts}.

%% file: 5_experiment/experiment.tex
\section{Experiments}

\subsection{Setup}
\textbf{Training.}
Following prior work~\cite{audsemthinker, audio_thinker, audio_mcq, omni_r1}, we adopt Qwen2.5-Omni-7B as our base model, the backbone shared by most recent audio reasoning models.
We apply our two-stage training as described in Section~\ref{sec:multi-stage}. While recent online-RL work for audio reasoning mainly focuses on improving GRPO training~\cite{audio_thinker, cesar}, we instead show that an LLT-trained model is itself a strong initialization that is readily compatible with online RL. We therefore additionally apply GRPO on AVQA~\cite{avqa}, an audio question answering dataset widely used in prior audio RL work~\cite{omni_r1, r1_aqa}. We randomly sample 8K questions and use a simple rule-based reward that assigns 1 if the answer inside the \texttt{<answer>} block is correct and 0 otherwise. More training details are given in Appendix~\ref{appen:training_details}.

\noindent\textbf{Evaluation Details.} We evaluate on three widely used audio reasoning benchmarks: MMAU Test-mini~\cite{mmau} and MMAR~\cite{mmar}, which cover general audio understanding and reasoning across speech, sound, and music, and MMSU~\cite{mmsu}, which focuses on speech perception and reasoning. More details are given in Appendix~\ref{appen:evaluation_details}.

\subsection{Results}
\label{sec:analysis}

\input{4_analysis/figure_bargraph}

\paragraph{Does multi-aspect dense audio description benefit audio reasoning?}
To isolate the effect of description richness (Section~\ref{sec:stage1}) on reasoning, we feed audio descriptions with varying levels of information to a strong text LLM (Qwen3-32B~\cite{qwen3} in thinking mode, used only for this analysis) and have it answer the questions from the descriptions alone. We evaluate on general audio understanding and reasoning benchmarks (MMAU Test-mini, MMAR, MMSU) as well as specialized benchmarks targeting low-level acoustic attributes (WoW-Bench~\cite{wow_bench}) and temporal reasoning (TREA~\cite{trea}).

Specifically, starting from the semantic description produced by Qwen3-Omni-Captioner, we progressively add acoustic and temporal information (Section~\ref{sec:stage1}) to form the full multi-aspect dense description. As shown in Figure~\ref{fig:analysis_stage1}, adding acoustic and temporal information yields consistent gains: each aspect benefits not only its corresponding specialized benchmark (WoW-Bench for acoustic, TREA for temporal) but also the general multi-task benchmarks. This result indicates that multi-aspect dense audio descriptions offer effective perceptual grounding and cues for generating high-quality reasoning traces in text-based LLMs.

\input{4_analysis/table_data_ablation}

\input{4_analysis/figure_cognitive_analysis}
\paragraph{Do incorrect reasoning traces benefit audio reasoning?}
Section~\ref{sec:stage4} introduces correct-to-refine traces $(r^+ \rightarrow r^{++})$ and incorrect-to-correct traces $(r^- \rightarrow r^{++})$ in addition to the original short CoT $r^+$. While the latter is intended to encourage reflective reasoning, exposing the model to incorrect intermediate steps $r^-$ may introduce noise during SFT. We compare three SFT settings under both a small-scale (50k QA) and a large-scale (full dataset) regime: $r^{+}$ only; $r^{+}$ with correct-to-refine $(r^{+}\!\rightarrow\! r^{++})$; and all three types, additionally including incorrect-to-correct $(r^{-}\!\rightarrow\! r^{++})$.

As shown in Table~\ref{tab:analysis_behavior}, adding correct-to-refine traces achieves the best task accuracy, while additionally including incorrect-to-correct traces $(r^- \rightarrow r^{++})$ degrades performance, especially at small scale. We attribute this result to the model's difficulty in distinguishing intermediate $r^-$ from $r^+$ under limited supervision, a gap that narrows with scale but is not fully resolved.

Analysis on model responses tells a different story. Quantifying cognitive behaviors on MMAR~\cite{mmar} using the framework of \citet{cognitive} and measuring response length (Figure~\ref{fig:cognitive_analysis}), we find that models trained without $(r^- \rightarrow r^{++})$ exhibit substantially fewer cognitive behaviors and shorter traces. This result reveals a fundamental observation that $(r^- \rightarrow r^{++})$ is essential to install cognitive behaviors such as verification and backtracking, but directly including it in SFT compromises accuracy.

This finding motivates the two-stage design in Section~\ref{sec:multi-stage}, where $(r^- \rightarrow r^{++})$ is reserved for DPO rather than SFT. Using only 60k preference pairs from the full LLT trace pool, the resulting LLT SFT~$\rightarrow$~DPO pipeline achieves both stronger task accuracy and more pronounced cognitive behaviors than any SFT setting alone, as shown in the last row in Table~\ref{tab:analysis_behavior} and Figure~\ref{fig:cognitive_analysis}.

We further verify that these cognitive behaviors translate into actual error recovery, not just surface-level reflection.
We identify MMAR and MMAU Test-mini responses from the two-stage model that contain one of the cognitive triggers used in Stage~4 (e.g., \textit{Let me listen again}, \textit{Wait}), and compare the answer the model had reached \textit{before} the cognitive trigger to its final answer after the subsequent continuation (Appendix~\ref{appen:cognitive_trigger}, Table~\ref{tab:cognitive_correction}). Across these triggered traces, 8.7\% flip from incorrect to correct (true correction), while only 4.9\% flip from correct to incorrect (over-correction), a statistically significant net correction of $+3.8$ percentage points (McNemar test~\cite{mcnemar1947note}, $p{<}0.001$). Moreover, this triggered subset is intrinsically harder: the SFT-only model is substantially less accurate on triggered questions than on non-triggered ones (MMAR: 58.9\% vs.\ 67.2\%; MMAU: 72.6\% vs.\ 81.4\%; Fisher's exact test, both $p{<}0.01$). These results confirm that the two-stage model invokes cognitive behaviors selectively on harder questions, and that these in-trace revisions recover the correct answer more often than they introduce errors.

\input{5_experiment/table_main_results}

\paragraph{Comparison to baselines.}
The results are given in Table~\ref{tab:perf_comparison_models}. Our two-stage LLT pipeline achieves substantial improvements over the Qwen2.5-Omni-7B baseline, with consistent gains across MMAU Test-mini, MMAR, and MMSU. Applying GRPO on AVQA~\cite{avqa}, we observe further improvements in MMAR and MMAU Test-mini. We additionally compare against LALMs of similar size (7B) and audio reasoning models using SFT or GRPO on AVQA-inclusive datasets. Our approach outperforms most of them while requiring only SFT and DPO, and performs comparably to those that use LLM-based~\cite{audio_thinker} or complex process rewards~\cite{cesar}. Moreover, compared to Omni-R1 (AVQA)~\cite{omni_r1} and Ke-Omni-R~\cite{ke-omni-r}, which share the same backbone of Qwen2.5-Omni-7B and outcome-based reward, LLT initialization yields stronger GRPO performance, indicating a stronger starting point for online RL.

\cready{
\input{5_experiment/table_ablation}

\noindent\textbf{Ablation on Incorrect Reasoning Traces (Stage~3).} As described in Section~\ref{sec:stage3}, we synthesize incorrect reasoning traces using $M_{\text{instruct}}$ rather than sampling them from the base LALM. A natural alternative is to prompt the base LALM on the source questions and retain outputs whose final answers are incorrect. We compare these two sources under two settings: (i)~a small-scale SFT setting that fine-tunes on 50k QA pairs using all three trace types ($r^{+}$, $r^{+}\!\rightarrow\! r^{++}$, and $r^{-}\!\rightarrow\! r^{++}$), and (ii)~our two-stage setting, where DPO is applied on top of the full-trace SFT model using preference pairs whose $r^{-}$ comes from the respective source. As shown in the second block of Table~\ref{tab:ablation}, synthesized traces generally outperform naturally occurring ones, indicating that the controlled, perceptually grounded errors produced by our pipeline provide a more effective corrective signal. More detailed analysis is given in Appendix~\ref{appen:negative_analysis}.

\noindent\textbf{Ablation on DPO Pair Selection.}
The two-stage pipeline (Section~\ref{sec:multi-stage}) combines three types of preference pairs.
To verify whether the incorrect-to-correct pair $(r^- \rightarrow r^{++} \succ r^-)$ genuinely contributes to the observed improvements, we ablate two variants. First, we remove the $(r^- \rightarrow r^{++} \succ r^-)$ pair, leaving only $(r^+ \succ r^-)$ and $(r^+ \rightarrow r^{++} \succ r^-)$, and increase each remaining pair type to 30k to match the original total scale. Second, we add a compactness-preference pair $(r^+ \succ r^+ \rightarrow r^{++})$ at 20k, following prior work~\cite{lpt, lgt} that uses such pairs to prefer shorter, compact reasoning when the model already reaches the correct answer.
Results are presented in the third block of Table~\ref{tab:ablation}. Including the $(r^- \rightarrow r^{++} \succ r^-)$ pair performs comparably on MMAU Test-mini while outperforming both variants on MMAR and MMSU, and adding the compactness pair $(r^+ \succ r^+ \rightarrow r^{++})$ does not help. This confirms that our DPO pair selection is well-suited for the two-stage pipeline.
}

\cready{
\input{5_experiment/table_generalize_model}

\noindent\textbf{Applying LLT Pipeline to a Different Dataset.}
To investigate whether the four-stage LLT pipeline generalizes beyond AF-Think, we apply the same pipeline to a different source dataset.
Specifically, we randomly sample 50k questions from AudioMCQ~\cite{audio_mcq} and apply Stage~1 and Stage~4 to construct corresponding $r^{++}$ traces for each $r^+$.
We then compare two training settings: (i) using the original $r^+$ alone, and (ii) using both $r^+$ and $r^+ \rightarrow r^{++}$, following the SFT setup of our two-stage training pipeline.
Results are presented in the fourth block of Table~\ref{tab:ablation}.
Adding the $r^+ \rightarrow r^{++}$ traces consistently improves performance over the $r^+$-only baseline, demonstrating that our pipeline generalizes across different source datasets.

\noindent\textbf{Applying LLT Dataset to Additional Models.}
To examine whether LLT generalizes beyond the primary backbone of Qwen2.5-Omni-7B, we fine-tune additional LALMs on a subset of the LLT dataset consisting of $r^+$ and $r^+ \rightarrow r^{++}$ reasoning traces corresponding to SFT examples derived from 50K QA pairs.
Specifically, we fine-tune Qwen2-Audio-Instruct~\cite{qwen2_audio}, which has been widely used in prior audio reasoning work~\cite{audio_cot, audio_reasoner, r1_aqa}, and Qwen3-Omni-Instruct~\cite{qwen3_omni}, a more recent and stronger LALM.
For Qwen3-Omni-Instruct, we apply LoRA~\cite{lora} with rank 64 and alpha 128 due to compute constraints.
The results are shown in Table~\ref{tab:generalize_model}.
Across benchmarks, both models trained with LLT show consistent improvements over their base counterparts, suggesting that LLT generalizes to other LALMs, including a stronger backbone such as Qwen3-Omni-Instruct.
Additional experimental results are provided in Appendix~\ref{appen:additional}.
}

\input{5_experiment/figure_qual}

\paragraph{Qualitative Examples.}
As shown in Figure~\ref{fig:qual}, compared to the model trained with AF-Think, which produces an incorrect answer, the model trained with LLT initially makes a mistake but subsequently revisits the audio, applies more deliberate perceptual analysis, and corrects its reasoning to arrive at the correct answer.
More examples are available on the project page\footnote{\url{https://jaeyeonkim99.github.io/llt}}.

%% file: 4_analysis/figure_bargraph.tex
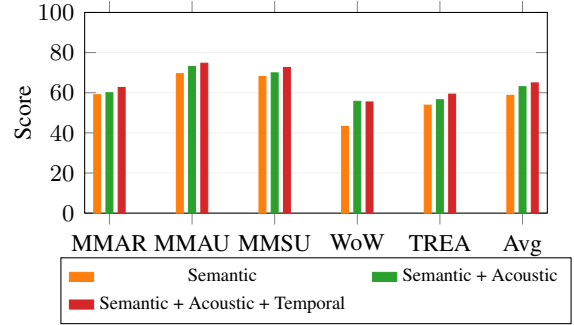
\begin{figure}[t]
\centering
\begin{tikzpicture}
\begin{axis}[
  ybar,
  width=\columnwidth,
  height=0.55\columnwidth,
  ymin=0, ymax=100,
  bar width=2.6pt,
  enlarge x limits=0.06,
  symbolic x coords={MMAR, MMAU, MMSU, WoW, TREA, Avg},
  xtick=data,
  tick label style={font=\footnotesize},
  label style={font=\footnotesize},
  ylabel={Score},
  ylabel style={
  font=\footnotesize,
  yshift=-5pt
  },
  ymajorgrids,
  grid style={opacity=.2},
  legend style={
    font=\scriptsize,
    at={(0.5,-0.22)},
    anchor=north,
    legend columns=2,
    /tikz/every even column/.append style={column sep=8pt}
  },
  legend image code/.code={
    \path[#1] (0cm,-0.06cm) rectangle (0.32cm,0.06cm);
  },
]

\addplot+[fill=taborange,draw=taborange] coordinates {(MMAR,59.0) (MMAU,69.5) (MMSU,68.1) (WoW,43.2) (TREA,53.8) (Avg,58.7)};
\addlegendentry{Semantic}

\addplot+[fill=tabgreen,draw=tabgreen] coordinates {(MMAR,60.0) (MMAU,73.1) (MMSU,69.9) (WoW,55.7) (TREA,56.5) (Avg,63.0)};
\addlegendentry{Semantic + Acoustic}

\addplot+[fill=tabred,draw=tabred] coordinates {(MMAR,62.6) (MMAU,74.7) (MMSU,72.6) (WoW,55.4) (TREA,59.3) (Avg,64.9)};
\addlegendentry{Semantic + Acoustic + Temporal}

\end{axis}
\end{tikzpicture}
\vspace{-2mm}
\caption{Accuracy of an LLM across different benchmarks using different types of audio descriptions. Semantic, Acoustic, and Temporal correspond to the respective components of the multi-aspect audio descriptions constructed in Stage 1 of LLT.}
\label{fig:analysis_stage1}
\vspace{-2mm}
\end{figure}

%% file: 4_analysis/table_data_ablation.tex
\begin{table}[t]
\centering

\resizebox{\linewidth}{!}{%
\begin{tabular}{lccc}
\hline
\textbf{Method} &
\makecell[c]{\textbf{MMAU} \\ \textbf{Test-mini}} &
\textbf{MMAR} &
\textbf{MMSU} \\
\hline
\multicolumn{4}{c}{\textbf{SFT with 50k QA Pairs}} \\
AF-Think ($r^{+}$) & 74.1 & 58.8 & 63.4 \\
LLT ($r^{+}$, $r^{+}\!\rightarrow\! r^{++}$) & \textbf{75.7} & \textbf{61.1} & \textbf{64.0} \\
LLT ($r^{+}$, $r^{+}\!\rightarrow\! r^{++}$, $r^{-}\!\rightarrow\! r^{++}$) & 74.6 & 60.2 & 63.3 \\
\hline
\multicolumn{4}{c}{\textbf{SFT with Full Reasoning Traces}} \\
AF-Think ($r^{+}$) & 73.9 & 59.9 & 63.7 \\
LLT ($r^{+}$, $r^{+}\!\rightarrow\! r^{++}$) & \textbf{76.3} & \textbf{62.2} & \textbf{65.0} \\
LLT ($r^{+}$, $r^{+}\!\rightarrow\! r^{++}$, $r^{-}\!\rightarrow\! r^{++}$) & 75.8 & 61.4 & 64.1 \\
\hline
\multicolumn{4}{c}{\textbf{Two-Stage Training}} \\
LLT (SFT $\rightarrow$ DPO) & \textbf{78.2} (+1.9) & \textbf{63.7} (+1.5) & \textbf{66.0} (+1.0) \\
\hline
\end{tabular}%
}
\caption{Comparison across different training setups. Notation for reasoning traces follows Section~\ref{sec:method}. Parentheses indicate the trace types used for SFT. The last block reports the full two-stage pipeline (SFT~$\rightarrow$~DPO). Best result in each block is bolded.}
\label{tab:analysis_behavior}
\end{table}

%% file: 4_analysis/figure_cognitive_analysis.tex
\begin{figure}[t]
    \centering
    \includegraphics[width=0.96\linewidth]{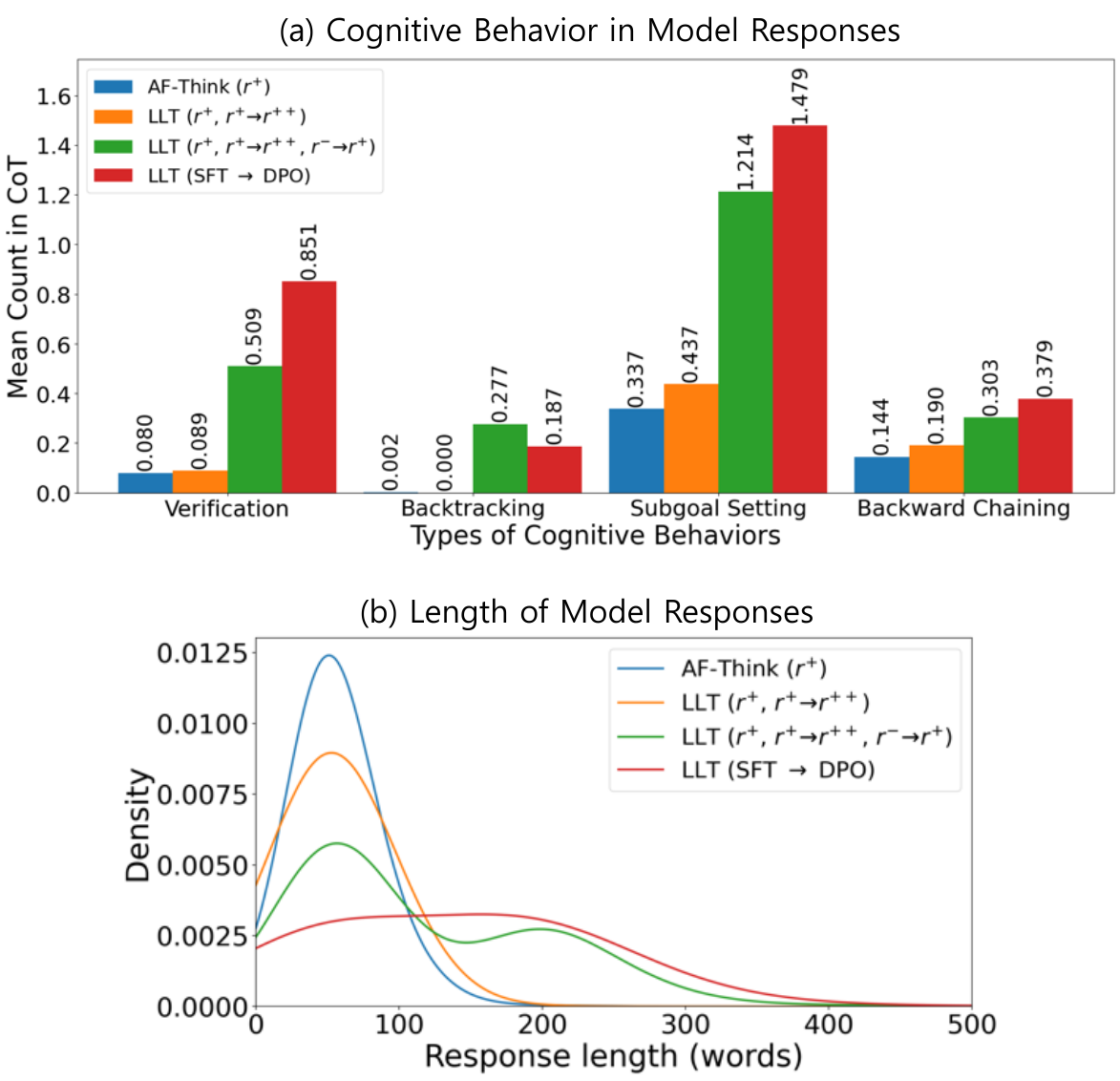}
    \vspace{-0.5\baselineskip}
    \caption{Cognitive analysis of model responses on MMAR.
(a) Mean occurrence of different cognitive behaviors (b) Distribution of response length (number of words).}
    \label{fig:cognitive_analysis}
\end{figure}

%% file: 5_experiment/table_main_results.tex
\begin{table}[t]
\centering

\resizebox{\linewidth}{!}{%
\begin{tabular}{lccc}
\hline
\textbf{Method} &
\makecell[c]{\textbf{MMAU} \\ \textbf{Test-mini}} &
\textbf{MMAR} &
\textbf{MMSU} \\
\hline
\multicolumn{4}{c}{\textbf{Large Audio Language Models}} \\
Qwen2-Audio-Instruct \citeyearpar{qwen2_audio} & 59.6 & 30.0 & 53.3\\
Kimi-Audio \citeyearpar{kimi_audio} & 68.2 & 57.6 & 59.3 \\
Audio Flamingo 3 \citeyearpar{af3} & 73.3 & 58.5 & 61.4 \\

\multicolumn{4}{c}{\textbf{Audio Reasoning Models}} \\
\multicolumn{4}{c}{\textit{SFT-based}} \\
Audio-Reasoner \citeyearpar{audio_reasoner} & 67.7 & 36.8 & 49.2 \\
AudSemThinker$^{\dagger\ddagger}$ \citeyearpar{audsemthinker} & 71.2 & 50.0 & 60.2 \\
\multicolumn{4}{c}{\textit{GRPO on AVQA-Inclusive sets}} \\
R1-AQA \citeyearpar{r1_aqa}  & 68.9 & 50.8 & 61.6 \\
Ke-Omni-R$^{\dagger}$ \citeyearpar{ke-omni-r} & 74.6 & 60.9 & 62.1 \\
Omni-R1$^{\dagger}$ \citeyearpar{omni_r1}  & 77.0 & 63.4 & --   \\
Omni-R1 (AVQA)$^{\dagger}$    &  75.8 & 61.2  & --     \\
CESAR$^{\dagger}$ \citeyearpar{cesar} & 77.1 & 62.7 & \textbf{64.2} \\
Audio-Thinker$^{\dagger}$ \citeyearpar{audio_thinker} & \textbf{77.7} & \textbf{65.3} & --   \\
\hline
\multicolumn{4}{c}{\textbf{Our Methods}} \\
Qwen2.5-Omni-7B      & 71.5 & 56.7 & 60.6 \\
+ LLT (SFT $\rightarrow$ DPO) & 78.2 & 63.7 & \textbf{66.0} \\
+ LLT + AVQA GRPO & \textbf{78.7} & \textbf{64.9} & 65.3\\
\hline
\end{tabular}%
}
\caption{Performance comparison across different models.  Omni-R1 (AVQA) indicates the variant trained on the naive AVQA dataset. For clarity, we highlight in bold the best-performing existing method and the best-performing model among our approaches. $\dagger$~denotes methods built on the same Qwen2.5-Omni-7B backbone as ours; $\ddagger$~denotes results reproduced by us.}
\label{tab:perf_comparison_models}
\end{table}

%% file: 5_experiment/table_ablation.tex
\begin{table}[t]
\centering
\resizebox{\linewidth}{!}{%
\begin{tabular}{lccc}
\hline
\textbf{Setting} &
\makecell[c]{\textbf{MMAU} \\ \textbf{Test-mini}} &
\textbf{MMAR} &
\textbf{MMSU} \\
\hline
Qwen2.5-Omni baseline & 71.5 & 56.7 & 60.6 \\
\hline
\multicolumn{4}{c}{\textbf{Stage 3: Source of Incorrect Traces}} \\
\multicolumn{4}{c}{\textit{SFT on 50k QA Pairs}} \\
 From base LALM & 72.9 & 58.3 & \textbf{64.4}  \\
 From $M_{\text{instruct}}$ & \textbf{74.6} &\textbf{60.2}  & 63.3 \\
\multicolumn{4}{c}{\textit{Two-Stage Training}} \\
From base LALM & 77.9 & \textbf{63.7} & 65.8  \\
From $M_{\text{instruct}}$  & \textbf{78.2} & \textbf{63.7} & \textbf{66.0} \\
\hline
\multicolumn{4}{c}{\textbf{DPO Pair Composition}} \\
LLT (SFT $\rightarrow$ DPO, full)              & 78.2 & \textbf{63.7} & \textbf{66.0} \\
\quad $-\;(r^{-}\!\rightarrow\! r^{++} \succ r^{-})$ &  78.3 &  62.2 &  64.4 \\
\quad $+\;(r^{+} \succ r^{+}\!\rightarrow\! r^{++})$ &  \textbf{78.4 }&  62.4 & 65.0 \\
\hline
\multicolumn{4}{c}{\textbf{Generalization to AudioMCQ~\cite{audio_mcq}}} \\
$r^{+}$ only                            &  73.2 & 57.7 & 63.8  \\
$r^{+}$, $r^{+}\!\rightarrow\! r^{++}$  &  \textbf{73.8} & \textbf{61.1} & \textbf{64.3} \\
\hline
\end{tabular}%
}
\caption{Ablation studies on the LLT pipeline. Each block isolates one design choice: the source of incorrect traces (Stage~3), DPO pair composition, and generalization to a different source dataset. The best result in each block is in bold.}
\label{tab:ablation}
\end{table}

%% file: 5_experiment/table_generalize_model.tex
\begin{table}[t]
\centering

\resizebox{0.9\linewidth}{!}{%
\begin{tabular}{lccc}
\hline
\textbf{Method} &
\makecell[c]{\textbf{MMAU} \\ \textbf{Test-mini}} &
\textbf{MMAR} &
\textbf{MMSU} \\
\hline
Qwen2-Audio-Instruct    &  59.6 & 30.0 & 53.3 \\
\quad + LLT SFT         &  \textbf{67.8} &  \textbf{53.8}& \textbf{59.0} \\
\hline
Qwen3-Omni-Instruct & 77.5  & 66.9  & 69.0 \\
\quad + LLT SFT & \textbf{79.6} & \textbf{67.9} & \textbf{71.6} \\
\hline

\hline
\end{tabular}%
}
\caption{Generalization of LLT to additional LALM backbones.}
\label{tab:generalize_model}
\end{table}

%% file: 5_experiment/figure_qual.tex
\begin{figure}[t]
    \centering
    \includegraphics[width=0.95\linewidth]{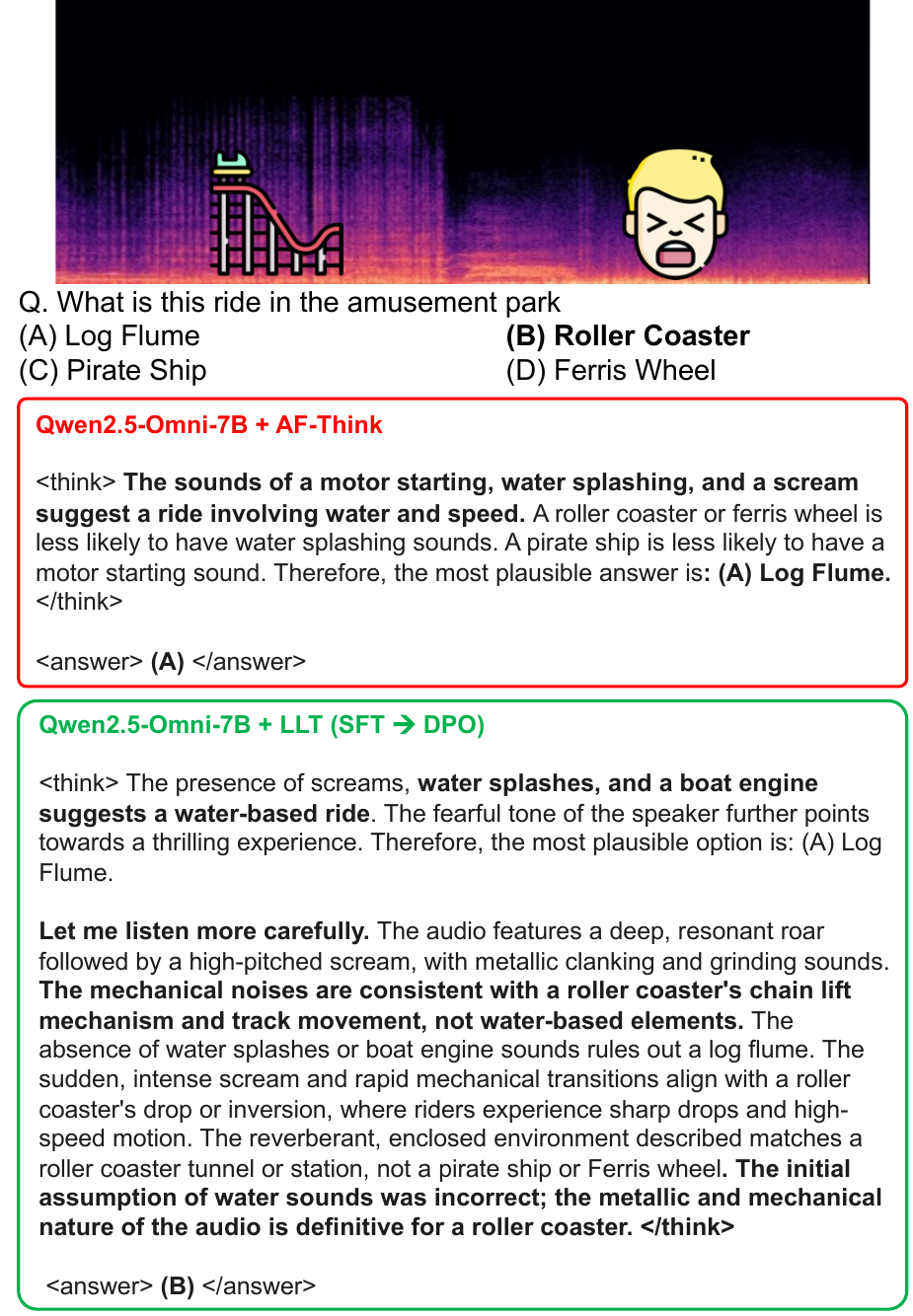}
    \vspace{-0.5\baselineskip}
    \caption{Qualitative comparison of responses from Qwen2.5-Omni-7B trained with AF-Think and with the proposed LLT on the same question.}
    \label{fig:qual}
\end{figure}

%% file: 7_appendix/appendix.tex
\section{Extended Related Work}
Recent large reasoning models \cite{jaech2024openai, deepseek} have demonstrated impressive performance on complex reasoning tasks such as mathematics and code generation by leveraging increased computation at inference time and producing long reasoning traces. Similar trends have also been observed in vision-centric reasoning tasks \cite{lpt, lgt}, where longer reasoning traces enable models to refine predictions through extended reasoning. In contrast, explicit strategies for test-time scaling in the audio reasoning domain remain largely underexplored.

Prior work indicates that effective test-time scaling is closely tied to structured cognitive behaviors, such as verification and backtracking \cite{cognitive, yeo2025demystifying}. In the visual domain, approaches such as Long Perceptual Thoughts \cite{lpt} and Long Grounded Thoughts \cite{lgt} attempt to encode these cognitive behaviors by distilling long reasoning traces from large reasoning models, leading to improved perceptual and grounded reasoning. Inspired by these efforts, we propose Long Listening Thoughts (LLT), which explicitly encodes cognitive behaviors for auditory reasoning. Unlike prior visual approaches that rely on detailed descriptions derived from relatively small-scale image datasets, LLT is designed for scalability by building upon existing large-scale audio CoT datasets. In addition, LLT incorporates a principled pipeline for constructing multi-aspect audio descriptions spanning semantic, acoustic, and temporal information. Finally, rather than directly distilling raw long reasoning traces, we adopt a thought-continuation strategy that more effectively teaches cognitive behaviors while producing concise and structured reasoning traces suitable for audio reasoning.

\section{Additional Experimental Results}
\label{appen:additional}
\input{7_appendix/table_correction}
\subsection{Cognitive Trigger Analysis}
\label{appen:cognitive_trigger}
To verify that cognitive behaviors learned through the two-stage pipeline lead to actual error recovery, we analyze MMAR and MMAU Test-mini responses from the LLT~SFT~$\rightarrow$~DPO model that contain a Stage~4 cognitive trigger (e.g., \textit{Let me listen again}, \textit{Wait}, \textit{Let me check again}). For each such response, we split the trace at its first cognitive trigger and prompt GPT-4.1-mini to identify the answer the model had committed to immediately before the trigger; responses for which no committed pre-trigger answer can be determined are excluded. We then compare this pre-trigger answer with the final answer after the continuation, pooling both benchmarks for a reliable significance test. Table~\ref{tab:cognitive_correction} reports the resulting pre-trigger~$\rightarrow$~post-trigger transitions over 1{,}132 determined responses. True corrections (99 cases, 8.7\%) outnumber over-corrections (56 cases, 4.9\%), yielding a statistically significant net gain of $+3.8$ percentage points on the triggered subset (McNemar test~\cite{mcnemar1947note}, $p{<}0.001$). The effect holds on each benchmark individually (MMAR: $+4.0$\,pp, $p{=}0.019$; MMAU: $+3.6$\,pp, $p{=}0.019$).

\subsection{Ablation on Thought Continuation (Stage 4)}
\label{appen:ablation}
\input{7_appendix/table_ablation_stage4}

We  compare the thought expansion approach used in prior work \cite{lpt, lgt}, which distills long reasoning traces from large reasoning models, with our proposed thought continuation strategy. Both are evaluated under an SFT setting trained with $r^{+}$ and $r^{+}\!\rightarrow\! r^{++}$, matching the SFT stage of our two-stage pipeline.
The results are in Table~\ref{tab:ablation_stage4}. Thought continuation consistently outperforms thought expansion, demonstrating that generating refined reasoning traces after the initial thinking phase is more effective than directly distilling raw long reasoning traces for audio reasoning.

\subsection{\cready{Analysis on Synthetic Incorrect Traces}}
\label{appen:negative_analysis}
\input{7_appendix/table_nll}
\input{7_appendix/table_nll_separation}

\cready{
We further analyze why the incorrect traces synthesized in Stage~3 (Section~\ref{sec:stage3}) are more effective than negatives sampled from the base LALM, considering two angles: coverage and likelihood.

\noindent\textbf{Coverage.} Model-sampled negatives are only obtainable when the base LALM both answers incorrectly and produces a reasoning trace. From 50k source questions, after running the full four-stage pipeline including verification, a single sampling run from the base LALM yields 4.1k $r^{-}\!\rightarrow\! r^{++}$ traces, whereas synthesis with $M_{\text{instruct}}$ yields 31.4k. A second sampling run from the base LALM adds 4.3k more, but 1.8k of these are duplicate questions that are already covered. Model-based sampling is therefore difficult to scale, and limits the diversity of the errors it can express.

\noindent\textbf{Likelihood analysis.} To understand the mechanism beyond coverage, we measure the per-token negative log-likelihood (NLL) of each trace type under the base LALM (Qwen2.5-Omni-7B) and under the SFT checkpoint that serves as the reference policy for DPO, on 4k paired samples.

The results are given in Table~\ref{tab:nll}, suggesting different explanations for the two training stages. For SFT, which starts from the base model, model-sampled $r^{-}$ is already highly likely (0.767), far below $r^{+}$ (2.173). Note that $r^{+}$ comes from an existing CoT dataset and is thus unfamiliar to the model. Including such traces therefore exposes the model to little new information and may reinforce its existing error patterns, whereas synthetic $r^{-}$ (2.510) is unfamiliar to the base model.

For DPO, one might expect synthetic negatives to lie outside the model's distribution, but in practice the gap is small. Since $r^{-}$ is generated to follow the format and length of $r^{+}$, its NLL drops substantially after SFT (2.510 $\rightarrow$ 1.948), while model-sampled $r^{-}$ moves in the opposite direction (0.767 $\rightarrow$ 1.615). The gap between the two pools narrows from 1.743 to 0.333 under the reference policy. Synthetic negatives are also more clearly separated from the chosen responses under the reference policy, as measured by $\text{NLL}(\text{rejected}) - \text{NLL}(\text{chosen})$, across all three preference pair types used in our DPO stage (Table~\ref{tab:nll_separation}). This separation alone does not explain the performance difference, but together with the coverage gap above it characterizes how the two pools differ as preference data.
}

\subsection{\cready{Generalization to Long-Form Audio}}
\label{appen:generalization_long}
\input{7_appendix/table_mmau_pro}

\cready{
To examine whether our improvements transfer to longer audio, we evaluate on MMAU-Pro~\cite{mmau_pro}, which includes recordings of up to 10 minutes with an average duration of about 2 minutes. Table~\ref{tab:mmau_pro} reports the results. Despite training exclusively on clips of up to 30 seconds, the gains are retained on the $>$30s subset, indicating that the improved audio reasoning from LLT transfers, at least to some extent, to longer recordings.
}

\subsection{\cready{Generalization to Specialized Audio Domains}}
\label{appen:generalization_domain}
\input{7_appendix/table_medmosaic}
\input{7_appendix/table_wow_bench}

\cready{
To investigate whether these improvements transfer beyond general audio benchmarks, we additionally evaluate on two specialized domains.
For the medical domain, we use the publicly released subset of MedMosaic~\cite{medmosaic} (3,982 multiple-choice instances), reported in Table~\ref{tab:medmosaic}. For bioacoustics, we use WoW-Bench~\cite{wow_bench}, which is based on marine mammal vocalizations, reported in Table~\ref{tab:wow_bench}. In both cases the trained models improve overall performance, showing that LLT helps, at least to some extent, not only on general benchmarks but also in specialized domains.
}

\subsection{\cready{Analysis on the Source of Gains}}
\label{appen:source_of_gains}
\input{7_appendix/table_audio_ablation}

\cready{
To further investigate the source of the gains, we evaluate our models under two conditions: removing the audio entirely, and replacing it with Gaussian noise, following prior work~\cite{mmau, omni_r1, audio_mcq}. All conditions are evaluated under an identical protocol for both models to isolate the effect of the audio input. Table~\ref{tab:audio_ablation} reports the results.

The results show two effects. First, without any audio, our model still outperforms the base model, indicating that text-side reasoning itself improves. Second, when the audio is replaced with Gaussian noise, our model degrades by a larger margin than the base model and even falls below its own no-audio score, suggesting that it actively grounds its reasoning in the acoustic evidence and is therefore misled when that evidence is corrupted. Together, these results indicate that the gains reflect both improved reasoning and strengthened perceptual grounding.
}

\section{Human Inspection of Reasoning Traces}
\label{app:human_eval}

To complement the automated LLM-based verification described in Section~\ref{sec:stage4}, we conduct a human inspection of the final reasoning trace pairs in our dataset.
This inspection assesses data fidelity along multiple axes including audio grounding, textual artifacts, and reasoning quality.

\subsection{Evaluation Protocol}
We randomly sample 96 trace pairs from the final dataset: 48 \emph{Type A} pairs $(r^+, r^+ \to r^{++})$ and 48 \emph{Type B} pairs $(r^-, r^- \to r^{++})$.
Twelve annotators, all graduate students with prior research experience in audio or multimodal machine learning, participated on a voluntary basis, each evaluating 16 pairs.
Each pair was independently evaluated by two annotators, yielding 192 trace-level annotations.

Annotators were instructed to listen to the audio before evaluating each pair.
The interface displayed the audio clip, the question with the correct answer marked, the initial reasoning trace (Left), and the continuation (Right).
The textual audio description was not shown to annotators.
The full guidelines and the labeling interface are shown in Figure~\ref{fig:annotation_guidelines} and Figure~\ref{fig:annotation_interface}, respectively.
The evaluation proceeds in two phases.

\input{7_appendix/figure_interface}

\paragraph{Question validity.}
Annotators first judge whether the question is valid: \emph{Is the question answerable given the audio?}
Yes if the question can be reasonably answered after listening, possibly with some external knowledge to interpret.
No if the question depends on audio events or attributes that are not present at all, making it impossible to answer in any grounded way or to meaningfully evaluate the reasoning.
If the question is judged invalid, the trace-level evaluation is skipped for that pair.

\paragraph{Per-trace evaluation.}
For pairs with valid questions, annotators independently evaluate the left and right traces along binary (Yes/No) criteria:

\begin{itemize}
    \item \textbf{Hallucination}: Does the trace mention audio events or attributes with no audible basis in the audio (e.g., claiming ``a dog barks'' when no dog-like sound is present)?
    \item \textbf{Textual artifact}: Does the trace explicitly reference information not derivable from listening to the audio alone, suggesting it was borrowed from an external textual description (e.g., ``the description mentions\ldots,'' ``according to the metadata\ldots,'' ``the caption says\ldots'')?
    \item \textbf{Reasoning valid}: Independent of the correct answer, is the reasoning logically coherent and plausible step-by-step?
\end{itemize}

For Type B pairs, the left trace ($r^-$) is by design incorrect, so \emph{hallucination} and \emph{reasoning valid} are replaced with a single judgment of \emph{plausibly incorrect}: whether the incorrect reasoning reflects a believable confusion (e.g., mishearing an ambiguous sound) rather than random nonsense unrelated to the audio.

\paragraph{Pair-level evaluation.}
Annotators additionally provide a pair-level judgment depending on the type:
\begin{itemize}
    \item \textbf{Type A: Better reasoning} (left / right) --- Which reasoning is better overall, considering audio grounding, detail, and reasoning quality?
    \item \textbf{Type B: Successful correction} (Yes/No) --- Does the continuation reconsider the audio and arrive at reasoning that points toward the correct answer?
\end{itemize}

\subsection{Results}

\input{7_appendix/table_human_eval}

The human inspection results are summarized in Table~\ref{tab:human_eval}

\noindent\textbf{Question validity.} 90\% of inspected questions were judged valid (88\% for Type~A, 92\% for Type~B).

\noindent\textbf{Trace quality (Type A).}
Among valid pairs, annotators preferred the continuation $r^{++}$ over the initial trace $r^{+}$ on overall reasoning quality in 65\% of cases (vs.\ 35\% for $r^{+}$; forced choice, no tie option).
Textual artifacts are rare ($\le$2\%) and reasoning is judged valid for nearly all traces (96\% for $r^{+}$, 99\% for $r^{++}$), indicating that continuation deepens reasoning while keeping it logically coherent.

\noindent\textbf{Correction quality (Type B).}
77\% of the incorrect traces $r^{-}$ were judged plausibly incorrect (realistic listener mistakes rather than random nonsense), supporting the design goal of Stage~3 for producing mistake-like errors.
Notably, the continuation successfully corrected the reasoning in 93\% of cases, providing direct human evidence that incorrect-to-correct traces $(r^{-}\!\to\! r^{++})$ reconsider the audio and arrive at the correct answer.

\noindent\textbf{Hallucination.}
Annotation-level hallucination rates are 13\% ($r^{+}$) to 18--19\% ($r^{++}$), indicating that the continuation does not substantially increase hallucination despite producing longer and more detailed reasoning traces.
Additionally, most flags come from a single annotator: when both annotators evaluate the same trace, they both flag hallucination in only 5\% of cases, while at least one flags it in 26\%.
This suggests that the majority of flags reflect borderline judgments rather than clear-cut hallucination.

\noindent\textbf{Inter-annotator agreement.}
We report raw agreement, Cohen's $\kappa$~\cite{cohen1960coefficient}, and Krippendorff's $\alpha$~\cite{krippendorff2011computing} in Table~\ref{tab:human_eval_iaa}.
Question validity and hallucination reach fair agreement ($\kappa{\approx}0.22$, 79--85\% raw agreement) under the \citet{landis1977measurement} scale.
Criteria with skewed label distributions (textual artifact, reasoning valid, successful correction) reach high raw agreement (88--97\%) but near-zero $\kappa$/$\alpha$ due to the prevalence paradox. We rely on raw agreement for these items.
The \emph{better-reasoning} preference shows genuine disagreement ($\kappa{=}{-}0.09$, 50\% raw agreement), reflecting the subjectivity of holistic quality judgments. We therefore treat the 65\% Type-A preference as an aggregate tendency rather than a consensus.

\section{Examples from the LLT pipeline}

\subsection{Acoustic Description Example}
\label{appen:acoustic_example}

\input{7_appendix/figure_spectrogram}

The three examples below show model outputs when they are prompted to describe detailed acoustic characteristics of the given audio such as frequency, duration, amplitude, and timbre. The audio is from \texttt{ESC-50/3-151081-B-20} (baby crying), and the corresponding spectrogram is shown in Figure~\ref{fig:appen_spectrogram}.

Qwen2-Audio-Instruct fails to follow the instruction and introduces unsupported acoustic claims.
Gemini-2.5-Pro~\cite{gemini_25} produces plausible-sounding but incorrect details, for example,
asserting rapid modulation that is absent in the signal.
In contrast, our Spectrogram+VLM approach yields the most accurate and grounded characterization
among the three.

\begin{figure*}[t]
\begin{promptbox}{Description from Qwen2-Audio-Instruct}
The audio features a male singing in English with a sad mood. The characteristics include a key of Bb minor, time signature of 4/4, and a tempo of 150.0 bpm.
\end{promptbox}
\caption{Acoustic description generated by Qwen2-Audio-Instruct.}
\label{fig:acoustic-qwen2audio}
\end{figure*}

\begin{figure*}[t]
\begin{promptbox}{Description from Gemini-2.5-Pro}
The audio is a 4-second snapshot of a complex vocal event. While short, it contains a significant amount of acoustic information that our brains immediately interpret as a specific emotional state: distress.
Here is a more detailed look at the components:

\medskip

\textbf{Frequency (Pitch)}

\medskip

Frequency corresponds to our perception of pitch. In this audio, the cry is characterized by:

\medskip

\textbf{High Fundamental Frequency:} The baseline pitch is high, which is a direct result of the small and thin vocal folds of an infant. Just as a thin guitar string produces a higher note than a thick one, an infant's smaller vocal anatomy produces a higher-pitched sound.

\medskip

\textbf{Wide and Rapid Fluctuation (Pitch Contour): The pitch is not steady. It sweeps up and down rapidly, creating the classic ``wailing'' sound.} This variation, or pitch contour, is a key acoustic cue for emotion. The initial sound starts at a certain pitch, then rises sharply, and falls again in a pattern that conveys urgency and discomfort. This dynamic shifting is far more attention-grabbing than a monotone sound would be...
\end{promptbox}
\caption{Acoustic description generated by Gemini-2.5-Pro.}
\label{fig:acoustic-gemini}
\end{figure*}

\begin{figure*}[t]
\begin{promptbox}{Description from Spectrogram + Qwen3-VL-32B}
\textbf{Duration:} The audio spans approximately 4.5 seconds, as indicated by the horizontal axis (time scale from 0 to 4.5). The signal is continuous with distinct segments: a sustained sound from ~0.0 to ~1.8 seconds, followed by a brief silence or low-energy period (~1.8--2.3 seconds), and then a return of similar sustained sound from ~2.3 to ~4.5 seconds.

\medskip

\textbf{Frequency:} The frequency range extends from 0 Hz to approximately 19,845 Hz, as shown on the vertical axis. The dominant energy is concentrated in the lower to mid-frequency range (below 11,025 Hz), particularly between 2,205 Hz and 6,615 Hz, where dense, periodic horizontal striations are visible -- indicative of harmonic structure. There are also consistent, weaker harmonics extending into the higher frequencies (up to ~15,435 Hz), especially during the first segment. \textbf{A few transient high-frequency components appear around 1.0--1.5 seconds at ~17,640--19,845 Hz, suggesting possible percussive or formant-like events.} Between 1.8--2.3 seconds, there is a noticeable drop in spectral content, indicating a pause or reduction in vocal/sonic activity...
\end{promptbox}
\caption{Acoustic description generated by our Spectrogram + Qwen3-VL-32B approach.}
\label{fig:acoustic-ours}
\end{figure*}

\subsection{Stage 1 Description Example}
\label{appen:description_example}
The following figures show the three components of the multi-aspect dense description generated by Stage~1 for an example audio clip (VGGSound/-0jeONf82dE\_000021): the semantic description (Figure~\ref{fig:stage1-semantic}), the acoustic description from spectrogram analysis (Figure~\ref{fig:stage1-acoustic}), and the temporal description aggregating outputs from multiple specialized models (Figure~\ref{fig:stage1-temporal}).

\begin{figure*}[t]
\begin{promptbox}{Stage 1 Example: Semantic Description (VGGSound/-0jeONf82dE\_000021)}
The audio opens with the unmistakable sound of a powerful V8 engine revving, its deep, throaty roar filling the space and immediately conjuring a sense of American muscle and automotive culture. Over this mechanical backdrop, a male voice---gruff, gravelly, and distinctly Southern---asks, ``Do you smell barbecue?'' His delivery is casual and relaxed, evoking the atmosphere of a Southern cookout or gathering. As the engine sound fades, a second male voice, bright and energetic, exclaims ``Hulk!'' with a playful, comic-book inflection, referencing the iconic Marvel character and introducing a whimsical, pop-culture element to the scene.

\medskip

The tone shifts as a third voice, calm and smooth, delivers the tagline ``Fresh, smooth, real. Bud Light.'' The voice is polished and resonant, set against a clean, studio-produced instrumental bed featuring a funky, syncopated bass line, crisp drum machine beat, and subtle electric guitar or keyboard chords. The music and voice are tightly mixed, with the jingle's rhythm and melody reinforcing the brand's identity and creating a memorable, catchy advertisement. This segment is unmistakably a commercial for Bud Light beer, crafted to evoke a sense of coolness and authenticity.

\medskip

Suddenly, the mood pivots again as a fourth voice, high-pitched and enthusiastic, exclaims, ``Cooler rocketslide!'' The delivery is reminiscent of a young boy or a character from a children's animated show, introducing a playful, imaginative twist. The music continues, but the tone is now more lighthearted and comedic, suggesting a family-friendly or youth-oriented context. The final voice, deeper and more formal, calmly states, ``It's all here,'' serving as a closing statement that ties the varied elements together and signals the end of the commercial.

\medskip

The audio clip is a tightly produced, high-fidelity advertisement that blends automotive, barbecue, comic-book, and pop-culture references with the recognizable Bud Light brand jingle. It features a sequence of distinct male voices---each with unique regional and stylistic traits---delivered in a montage that moves from Southern barbecue to Marvel fandom, from beer commercial to playful children's exclamation. The music shifts from raw rock to funky pop, maintaining crisp, professional sound quality throughout.
\end{promptbox}
\caption{Stage~1 example: semantic description generated by Qwen3-Omni-Captioner.}
\label{fig:stage1-semantic}
\end{figure*}

\begin{figure*}[t]
\begin{promptbox}{Stage 1 Example: Acoustic Description (VGGSound/-0jeONf82dE\_000021)}
\textbf{Overview:} The spectrogram displays a linear frequency scale from 0 to 22,050~Hz and a time span of 0 to 9.01~seconds. The intensity is represented by a color gradient ranging from black ($-80$~dB) to yellow (0~dB).

\medskip

\textbf{Frequency:} Energy is concentrated primarily between 0 and 15,000~Hz, with the most intense components occurring between 2,500 and 12,500~Hz. Below 2,500~Hz, energy is present but less intense. Above 15,000~Hz, activity is sparse and appears as intermittent, low-intensity vertical streaks.

\medskip

\textbf{Amplitude/Intensity:} The highest intensities are observed in the mid-frequency range (approximately 2,500--12,500~Hz) during periodic time intervals, reaching levels around $-10$ to 0~dB. Between these peaks, the signal drops to around $-40$ to $-60$~dB, with background noise near $-70$ to $-80$~dB.

\medskip

\textbf{Timbre:} The signal exhibits a complex, broadband timbre with a strong presence of harmonic-like structures, particularly evident in the form of vertical striations and clustered bands.

\medskip

\textbf{Duration of events:} Acoustic activity occurs in recurring, short-duration bursts that last approximately 0.5 to 1.0~seconds each, separated by brief gaps of lower intensity, appearing roughly every 1.0 to 1.5~seconds.

\medskip

\textbf{Onset/Offset Characteristics:} Each burst exhibits a rapid onset, typically within 0.1--0.2~seconds, marked by a sudden increase in intensity across a broad frequency band. Offsets are similarly abrupt.

\medskip

\textbf{Spectral Characteristics:} The spectrum features distinct vertical striations and horizontal banding in the 2,500--12,500~Hz range, indicating a quasi-periodic structure with possible harmonic content.

\medskip

\textbf{Spatial/Reverberation:} No clear reverberant tails or echo patterns are visible, indicating minimal room reverberation or a direct, close-field recording condition.

\medskip

\textbf{Periodicity:} There is a clear periodicity in the temporal domain, with bursts occurring at approximately 1.0--1.5~second intervals throughout the recording.

\medskip

\textbf{Noise Components:} Background noise is relatively low, with large portions of the spectrogram near the $-70$ to $-80$~dB level.

\medskip

\textbf{Summary:} The signal consists of a series of short, high-intensity, broadband bursts occurring at regular intervals over 9.01~seconds, with a complex, harmonically rich spectral structure concentrated between 2,500 and 12,500~Hz, suggesting a modulated or pulsed acoustic source with minimal reverberation.
\end{promptbox}
\caption{Stage~1 example: acoustic description generated by Qwen3-VL-32B from the spectrogram.}
\label{fig:stage1-acoustic}
\end{figure*}

\begin{figure*}[t]
\begin{promptbox}{Stage 1 Example: Temporal Description (VGGSound/-0jeONf82dE\_000021)}
\textbf{Sound Event Detection (Threshold 0.2):}\\
\ \ [0.00--0.84\,s] Neigh, whinny \\
\ \ [0.76--1.40\,s] Background noise \\
\ \ [0.80--2.96\,s] Male speech, man speaking \\
\ \ [1.24--2.40\,s] Wind \\
\ \ [2.08--2.32\,s] Background noise \\
\ \ [2.76--3.00\,s] Wind \\
\ \ [3.04--9.01\,s] Music \\
\ \ [3.36--7.44\,s] Male speech, man speaking \\
\ \ [7.28--8.68\,s] Neigh, whinny \\
\ \ [8.24--9.01\,s] Male speech, man speaking

\medskip

\textbf{Sound Event Detection (Threshold 0.5):}\\
\ \ [0.20--0.52\,s] Neigh, whinny \\
\ \ [0.88--2.80\,s] Male speech, man speaking \\
\ \ [3.12--9.00\,s] Music \\
\ \ [3.44--5.88\,s] Male speech, man speaking \\
\ \ [6.16--7.32\,s] Male speech, man speaking \\
\ \ [7.40--7.96\,s] Neigh, whinny \\
\ \ [8.32--9.01\,s] Male speech, man speaking

\medskip

\textbf{Speech Transcription with Timestamps:}\\
\ \ [0.50--1.86\,s] ``Do you smell barbecue?''\\
\ \ [2.34--2.82\,s] ``Oh!''\\
\ \ [3.34--5.86\,s] ``Fresh, smooth, real, Bud Light.''\\
\ \ [6.14--7.38\,s] ``Cool a rocket sleigh.''\\
\ \ [8.22--9.00\,s] ``It's all here.''

\medskip

\textbf{Speaker Diarization:}\\
\ \ [0.44--3.12\,s] SPEAKER\_00 \\
\ \ [0.84--1.01\,s] SPEAKER\_01 \\
\ \ [3.41--9.09\,s] SPEAKER\_00

\medskip

\textbf{Music Structure Analysis:}\\
\ \ Tempo (BPM): 128.0\\
\ \ Beats Count: 19, Downbeats Count: 5\\
\ \ Beat Start Time: 0.75\,s, Downbeat Start Time: 0.75\,s\\
\ \ Average Beat Interval: 0.461\,s, Std Dev: 0.028\,s\\
\ \ Beat Stability: steady\\
\ \ Music Sections: [0.0--10.0\,s] verse

\medskip

\textbf{Chord Recognition:}\\
\ \ [2.93--5.94\,s] A:min
\end{promptbox}
\caption{Stage~1 example: temporal description aggregating timestamps from Whisper, pyannote, BEATs-based sound event detection, allin1 music structure analysis, and autochord chord recognition.}
\label{fig:stage1-temporal}
\end{figure*}

\subsection{Thought Expansion vs. Thought Continuation}
\label{appen:stage4_example}

The following reasoning traces correspond to the same example question,
\texttt{YouTube8M/iBH70m\_mjJw\_143\_10}, where \texttt{(D) Drums} is the correct answer.
Although both approaches arrive at the correct prediction, Thought Expansion repeatedly
restates the audio description and follows a more circuitous reasoning pattern.
In contrast, Thought Continuation directly references  auditory cues and produces
a more focused and refined rationale.

\begin{figure*}[t]
\begin{promptbox}{Example Question and Initial Reasoning Trace}
\textbf{Question}\\
Which instrument provides the rhythmic foundation in the audio? Choose one among the following options:\\
(A) Harmonica \\
(B) Flute \\
(C) Violin \\
\textbf{(D) Drums} \\
Please think and reason about the input audio before you respond.

\medskip

\textbf{Initial Incorrect CoT Trace} \\
The audio features a persistent, pulsing low-frequency tone that resembles a bowed string instrument's sustain, and the rhythmic chant aligns closely with the vibrato-like modulation in the bassline. While drums are present, they're masked by heavy compression. The violin's melodic contour, though subtle, drives the groove with syncopated bowing patterns, making it the true rhythmic anchor.
\end{promptbox}
\caption{Example question and initial incorrect CoT trace used to compare thought expansion and thought continuation. Correct answer: (D) Drums.}
\label{fig:thought-question}
\end{figure*}

\begin{figure*}[t]
\begin{promptbox}{Example of Thought Expansion}
\textbf{Let me listen again.} \textbf{The spectrogram shows} strong energy between 0 and 15 kHz with no harmonic peaks. The timbre is broadband and noise-like. The temporal description shows music and male singing throughout, but no specific mention of a violin. The acoustic description says no harmonic structure or periodicity. \textbf{The semantic description says} ``powerful, four-on-the-floor kick drum, sharp synthetic snare hits, and a pulsating, low-frequency bassline that forms the core of the groove.'' The question is about the rhythmic foundation. The rhythmic foundation is typically the kick drum in dance music. The bassline supports it, but the kick drum provides the steady pulse. Let me check the options: Harmonica, Flute, Violin, Drums.\textbf{ The semantic description clearly states }``powerful, four-on-the-floor kick drum, sharp synthetic snare hits, and a pulsating, low-frequency bassline that forms the core of the groove.'' So the kick drum is part of the drums. The bassline is also part of the rhythmic foundation, but the drums (kick and snare) are the primary rhythmic elements. The bassline is more of a melodic and harmonic element, but it also has rhythmic aspects. However, in dance music, the kick drum is the main rhythmic foundation.\textbf{ The acoustic description says:} ``Energy is distributed across the entire frequency range, but the most intense components are concentrated between 0 and 15,000 Hz.'' The low-frequency energy is from the kick and bass.\textbf{ The temporal description shows} music and male singing, but not specifically drums.\textbf{ But the semantic description explicitly says} ``powerful, four-on-the-floor kick drum, sharp synthetic snare hits'' which are drums. The question is ``which instrument provides the rhythmic foundation''. In electronic dance music, the rhythmic foundation is typically the kick drum. The options are Harmonica, Flute, Violin, Drums. Drums would include the kick and snare. The bassline is not an instrument in the options. So Drums would be the correct choice. The other options are all melodic instruments: harmonica, flute, violin. They are not typically used for rhythmic foundation in this context. The acoustic description says ``no harmonic structure'' which would suggest it's not a melodic instrument. Drums are percussive and non-harmonic. So the rhythmic foundation must be drums. So the answer is (D) Drums.

\medskip

\textless answer\textgreater\ (D)\ \textless/answer\textgreater
\end{promptbox}
\caption{Example of thought expansion: $M_{\text{reason}}$ repeatedly restates the audio description and follows a circuitous reasoning pattern.}
\label{fig:thought-expansion}
\end{figure*}

\begin{figure*}[t]
\begin{promptbox}{Example of Thought Continuation}
\textbf{Hmm.} The audio features a consistent, driving beat with a prominent kick drum pattern that pulses on every beat, forming the backbone of the rhythm. The high-energy electronic percussion includes sharp snare hits and steady hi-hats, but the most dominant rhythmic element is the deep, repetitive kick drum that anchors the entire track. There are no melodic or harmonic instruments like harmonica, flute, or violin present---the sound is purely synthetic and percussive, with no acoustic string or wind instruments detectable. The bassline, while low-frequency, is part of the electronic production and not a violin or other traditional instrument. \textbf{The drums, specifically the kick drum, are unmistakably the rhythmic foundation}.

\medskip

\textless answer\textgreater\ (D)\ \textless/answer\textgreater
\end{promptbox}
\caption{Example of thought continuation: $M_{\text{reason}}$ directly references auditory cues and produces a more focused, refined rationale.}
\label{fig:thought-continuation}
\end{figure*}

\section{Prompts for LLT}
\label{appen:prompts}

This section provides the full prompts used at each stage of the LLT pipeline. For Stage~1, Figure~\ref{fig:prompt-acoustic} shows the prompt given to the vision-language model to produce the acoustic description from the spectrogram. For Stage~2, Figure~\ref{fig:prompt-qgen} shows the prompt used to generate a more challenging question, and Figure~\ref{fig:prompt-qverify} the prompt used to verify the generated question, reasoning, and answer. For Stage~3, Figure~\ref{fig:prompt-wrongcot} shows the prompt used to synthesize an incorrect reasoning trace $r^-$. For Stage~4, Figure~\ref{fig:prompt-continuation} shows the thought-continuation prompt, and Figure~\ref{fig:prompt-verify-long} the prompt used to verify and minimally repair the resulting long reasoning traces. All prompts are used with the open-source models described in Section~\ref{sec:impl}.

\begin{figure*}[t]
\begin{promptbox}{Prompt for Acoustic Description (Stage~1)}
\textbf{System Prompt:}

You are an expert in acoustics and audio signal analysis.
You analyze spectrograms with high precision based on the visual evidence shown in the image.
Your descriptions must be concise, technically accurate, and fully aligned with the user's instructions.
You must follow the required output format exactly.

\medskip
\textbf{User Prompt:}

Please provide an acoustic description of the given spectrogram.
Describe key features such as frequency, amplitude/intensity, timbre, and duration of events.
When relevant, you may also comment on additional characteristics such as reverberation, periodicity,
spectral features, onset/offset characteristics, harmonic structure, and noise components.
When using numeric values, ensure they are reported accurately according to the axis labels and grid lines.
Do not infer or suggest what the audio event represents.

If a particular characteristic is not clearly observable in the spectrogram,
you can leave the corresponding field as an empty string (``'').

Return your response in the following JSON format, following the structure and level of detail illustrated in the example below.

\medskip
\{\\
\hspace{2em}"Overview": "", "Frequency": "", "Amplitude/Intensity": "", "Timbre": "",\\
\hspace{2em}"Duration of events": "", "Onset/Offset Characteristics": "",\\
\hspace{2em}"Spectral Characteristics": "", "Spatial/Reverberation": "",\\
\hspace{2em}"Periodicity": "", "Noise Components": "", "Others": "", "Summary": ""\\
\}

\medskip
Example:\\
\{\\
\hspace{2em}"Overview": "The spectrogram displays a linear frequency scale from 0 to 24,000 Hz and a time span of 0 to 18 seconds. The intensity is represented by a color gradient ranging from black ($-80$ dB) to yellow (0 dB).",\\
\hspace{2em}"Frequency": "Prominent energy is concentrated below 10,000 Hz, with the most intense components occurring between 0 and 5,000 Hz.",\\
\hspace{2em}"Amplitude/Intensity": "The highest intensities are localized primarily in the lower-frequency range below 3,000 Hz and occur in discrete time intervals.",\\
\hspace{2em}"Timbre": "Complex, predominantly broadband spectral texture with irregular energy distributions.",\\
\hspace{2em}"Duration of events": "Approximately 18 seconds, with intermittent short bursts of activity.",\\
\hspace{2em}"Onset/Offset Characteristics": "Events exhibit rapid onsets and offsets with sharp transitions.",\\
\hspace{2em}"Spectral Characteristics": "Dominated by broadband energy with irregular spectral shapes.",\\
\hspace{2em}"Spatial/Reverberation": "No clear reverberant tails or echo patterns are visible.",\\
\hspace{2em}"Periodicity": "No stable periodic structure is observed.",\\
\hspace{2em}"Noise Components": "Background noise remains near the noise floor.",\\
\hspace{2em}"Others": "", "Summary": "Short-duration broadband pulses dominate the signal."\\
\}
\end{promptbox}
\caption{Stage~1 prompt used with Qwen3-VL-32B to generate the acoustic description from the spectrogram image.}
\label{fig:prompt-acoustic}
\end{figure*}

\begin{figure*}[t]
\begin{promptbox}{Prompt for Question Generation (Stage~2)}
\textbf{System Prompt:}

You are a question generation assistant. Your task is to generate more challenging questions based on audio and original questions. You must strictly follow the user's instructions and output format. Return ONLY valid JSON and nothing else.

\medskip
\textbf{User Prompt:}

You are tasked with generating a MORE CHALLENGING question based on the given audio description and the original question.

\textless original\_question\textgreater\, \_\_ORIGINAL\_QUESTION\_\_ \textless /original\_question\textgreater\\
\textless original\_response\textgreater\, \_\_ORIGINAL\_RESPONSE\_\_ \textless /original\_response\textgreater\\
\textless audio\_description\textgreater\, \_\_AUDIO\_DESCRIPTION\_\_ \textless /audio\_description\textgreater

\medskip
Guidelines:
\begin{itemize}
\item The respondent will be given only the audio and the question. Do NOT explicitly mention or refer to the existence of any description or metadata. The reasoning should sound as if it comes solely from listening to the audio.
\item Preserve the same question format as the original whenever possible (multiple-choice or free-form).
\item Increase difficulty along one or both of the following dimensions:
\begin{itemize}
\item More audio-aware: requires closer, more precise listening.
\item More complex reasoning: requires additional reasoning steps (e.g., combining cues, resolving ambiguity, eliminating plausible alternatives).
\end{itemize}
\item Prefer to retain the core skill(s) tested by the original question. If preserving the same skill would force an unnatural question, design a new well-grounded question based on the same audio.
\item The question must be answerable from the audio alone.
\item For multiple-choice: create plausible distractors similar in difficulty and surface plausibility, while ensuring the correct answer can still be determined through careful listening.
\item Along with the question, generate (i) a short chain-of-thought (max 100 words) leading to the correct answer, and (ii) the final answer.
\end{itemize}

\medskip
Output: Return your output strictly in the following JSON format:\\
\{\\
\hspace{2em}"question": "\textless generated question\textgreater",\\
\hspace{2em}"response": "\textless short reasoning trace leading to the answer\textgreater",\\
\hspace{2em}"answer": "\textless choice letter for MCQ, or concise text for free-form\textgreater",\\
\hspace{2em}"choices": null or ["choice A", "choice B", "choice C", "choice D", ...]\\
\}
\end{promptbox}
\caption{Stage~2 prompt used with $M_{\text{instruct}}$ to generate a more challenging question from the original QA pair and the dense audio description.}
\label{fig:prompt-qgen}
\end{figure*}

\begin{figure*}[t]
\begin{promptbox}{Prompt for Question Verification (Stage~2)}
\textbf{System Prompt:}

You are a verification assistant. Your role is to evaluate the provided generated question according to the given verification criteria and produce a JSON judgment in the specified output format.

\medskip
\textbf{User Prompt:}

You are a verifier for audio-based questions. You will be given (i) an audio description (used only as a proxy for the audio), (ii) a question generated based on the audio, (iii) a model-generated reasoning trace, and (iv) a final answer.

\medskip
Your task is to verify whether the generated question and response satisfy the verification criteria below. Treat the audio description strictly as a proxy for the audio. Your judgment should reflect whether the question can be reasonably answered based on the given audio.

\medskip
\textbf{Verification Criteria}

\medskip
\textbf{1. Audio-based Solvability.} Determine whether the question is grounded in the audio signal and can be answered using audible evidence. A question is NOT audio solvable if:
\begin{itemize}
\item The required distinction is ambiguous or indistinguishable from the audio signal alone.
\item The question relies on hallucinated events/attributes that cannot be identified by listening.
\item The question relies on over-interpretation of the audio description.
\item The question depends on attributes not defined by the audio (e.g., visual details, metadata).
\item For multiple-choice: distractors cannot be reliably distinguished, or more than one choice remains equally plausible.
\end{itemize}

\textbf{2. Answer Correctness.} For multiple-choice, the selected answer must be correct relative to the provided choices. For non-multiple-choice, the free-form answer must be correct.

\textbf{3. Reasoning Validity.} The reasoning must be logically coherent, plausibly justify the final answer based on audible evidence, and must not mention or rely on the audio description or metadata. Minor stylistic issues are acceptable.

\medskip
\textbf{Output Format:} Return your judgment strictly in the following JSON format:\\
\{\\
\hspace{2em}"audio\_based\_solvable": true or false,\\
\hspace{2em}"answer\_correct": true or false,\\
\hspace{2em}"reasoning\_valid": true or false\\
\}

\medskip
\textbf{Input}\\
\textless audio\_description\textgreater\, \_\_AUDIO\_DESCRIPTION\_\_ \textless /audio\_description\textgreater\\
\textless question\textgreater\, \_\_QUESTION\_\_ \textless /question\textgreater\\
\textless reasoning\textgreater\, \_\_REASONING\_\_ \textless /reasoning\textgreater\\
\textless answer\textgreater\, \_\_ANSWER\_\_ \textless /answer\textgreater
\end{promptbox}
\caption{Stage~2 prompt used with $M_{\text{instruct}}$ to verify the generated question, reasoning, and answer.}
\label{fig:prompt-qverify}
\end{figure*}

\begin{figure*}[t]
\begin{promptbox}{Prompt for Incorrect CoT Generation (Stage~3)}
\textbf{System Prompt:}

You are an expert reasoning assistant. Your job is to produce coherent, logically structured reasoning traces according to the user's instructions. You must follow the required output format precisely.

\medskip
\textbf{User Prompt:}

\textless question\textgreater\, \_\_QUESTION\_\_ \textless /question\textgreater\\
\textless original\_cot\textgreater\, \_\_ORIGINAL\_COT\_\_ \textless /original\_cot\textgreater\\
\textless audio\_description\textgreater\, \_\_AUDIO\_DESCRIPTION\_\_ \textless /audio\_description\textgreater

\medskip
Above is a question and its original chain-of-thought (CoT) response that leads to the correct answer. Additional audio descriptions are also provided above. You may refer to these descriptions if needed when generating the reasoning trace.

\medskip
Your task is to generate an alternative reasoning trace that sounds logical and plausible but leads to an incorrect final answer.

\medskip
Requirements for the incorrect reasoning trace:
\begin{itemize}
\item It must be coherent, step-by-step, and internally consistent.
\item It should appear reasonable on the surface but include subtle mistakes, incorrect assumptions, misinterpretations of the audio (e.g., misperceiving timing or events), or overlooked details.
\item It must NOT contradict facts randomly; the reasoning should feel believable.
\item The incorrect answer must be a valid option for the question if it is multiple-choice.
\item Do NOT copy the original chain-of-thought. Produce a new and distinct reasoning process.
\item Do NOT change the wording of the question or the given choices.
\item The length should be similar to the original chain-of-thought and must not exceed \textasciitilde100 words.
\end{itemize}

Answer-format requirements: For multiple-choice with explicit letters, use ONLY the choice letter. For multiple-choice without letters, use the exact choice text. For free-form, provide a concise free-form answer.

\medskip
Return your output strictly in the following JSON format:\\
\{\\
\hspace{2em}"question": "\textless insert question\textgreater",\\
\hspace{2em}"original\_answer": "\textless insert original correct answer\textgreater",\\
\hspace{2em}"wrong\_answer": "\textless insert incorrect final answer\textgreater",\\
\hspace{2em}"wrong\_response": "\textless insert step-by-step reasoning leading to the incorrect answer\textgreater"\\
\}
\end{promptbox}
\caption{Stage~3 prompt used with $M_{\text{instruct}}$ to synthesize an incorrect reasoning trace $r^-$ from the original $(q, r^+, a^+)$ and the dense audio description.}
\label{fig:prompt-wrongcot}
\end{figure*}

\begin{figure*}[t]
\begin{promptbox}{Prompt for Thought Continuation (Stage~4)}
\textbf{System Prompt:}

You are an assistant whose role is to improve and correct reasoning.

\medskip
\textbf{User Prompt:}

\textless audio\_description\textgreater\, \_\_AUDIO\_DESCRIPTION\_\_ \textless /audio\_description\textgreater

\medskip
Answer the question below as if you are listening directly to the audio depicted by the description above. You may refer to that description for accuracy, but you must not explicitly mention or quote it. Avoid statements like ``The description mentions\ldots'' inside your response.

\medskip
You will be given a previously generated chain-of-thought inside a \textless reasoning\textgreater{} block. Continue the reasoning to solve the question accurately. You may correct any errors, incorrect assumptions, or flaws in the reasoning if they exist. Continue exactly from the last word of the given chain-of-thought, as if you are completing the unfinished sentence.

\medskip
Your continuation should:
\begin{itemize}
\item follow naturally from the text already written.
\item always extend the reasoning, even if you believe the original final answer is already correct.
\item use the following cognitive behaviors when helpful: \textit{verification} (systematically checking intermediate results), \textit{backtracking} (explicitly revising approaches when errors are detected), \textit{subgoal setting} (breaking a complex problem into manageable steps), and \textit{backward chaining} (reasoning backward from a desired outcome to the initial inputs).
\item optionally revise the final answer based on your improved reasoning. The answer may remain the same or change.
\item NOT mention or refer to the description explicitly; the reasoning must sound as if it comes solely from listening to the audio.
\end{itemize}

Place the final answer inside \textless answer\textgreater{}\ldots\textless /answer\textgreater{} after the reasoning.

\medskip
\textbf{Answer format:} For multiple-choice with explicit letters, use ONLY the letter, e.g., \textless answer\textgreater{} (B) \textless /answer\textgreater{}. For multiple-choice without letters, use the exact choice text, e.g., \textless answer\textgreater{} Grief \textless /answer\textgreater{}. For non-multiple-choice, use a concise answer, e.g., \textless answer\textgreater{} To achieve political goals by violent means \textless /answer\textgreater{}.

\medskip
\textless question\textgreater\, \_\_QUESTION\_\_ \textless /question\textgreater
\end{promptbox}
\caption{Stage~4 prompt used with $M_{\text{reason}}$ to continue an initial reasoning trace $r \oplus c$ with explicit cognitive behaviors.}
\label{fig:prompt-continuation}
\end{figure*}

\begin{figure*}[t]
\begin{promptbox}{Prompt for Long Reasoning Trace Verification (Stage~4)}
\textbf{User Prompt:}

You are a verifier for reasoning traces on audio-based questions. You will be given (i) an audio description (used only as a proxy for the audio), (ii) a question, (iii) the correct answer, (iv) a previous reasoning (context), and (v) a continued reasoning generated by the model that is intended to follow the previous reasoning.

\medskip
IMPORTANT: Evaluate ONLY the \textless continued\_reasoning\textgreater{}. The \textless previous\_reasoning\textgreater{} is provided only as context and may be correct or incorrect; do NOT judge or correct it. Your role is strictly to verify, not to rewrite or improve the content.

\medskip
\textbf{Verification Criteria}

\medskip
\textbf{1. Answer Correctness.} The final answer in the continued reasoning must match the correct answer. For free-form, preserve the core meaning (differences in specificity or phrasing are acceptable; do NOT rewrite the \textless answer\textgreater{} to match the correct answer verbatim). For multiple-choice with explicit letters, the choice letter must be correct; without letters, the selected text must match the correct choice.

\textbf{2. Reasoning Validity.} The continued reasoning must be logically coherent and consistent and plausibly support the final answer. Clear logical contradictions, hallucinated claims, or unsupported reasoning steps invalidate the reasoning.

\textbf{3. Format Validity.}
\begin{itemize}
\item \textit{Description / Metadata Mentions:} Must NOT explicitly mention the description or metadata (e.g., ``the description says\ldots'', ``according to the description\ldots''). The reasoning must sound as if it comes solely from listening to the audio.
\item \textit{Answer Tag Rules:} The final answer must appear inside the final \textless answer\textgreater{}\ldots\textless /answer\textgreater{} tag, and the tag must appear only once.
\item \textit{Answer Format Rules:} For free-form, preserve the original answer if correct. For MCQ with letters: \textless answer\textgreater{} (letter) \textless /answer\textgreater{}. Without letters: \textless answer\textgreater{} choice text \textless /answer\textgreater{}. For non-MCQ: \textless answer\textgreater{} concise answer \textless /answer\textgreater{}.
\item \textit{Forbidden Tags:} No other tags (e.g., \textless reason\textgreater{}, \textless think\textgreater{}, \textless reasoning\textgreater{}) should appear in the continued reasoning.
\end{itemize}

\textbf{Format Validation Policy:} If violations are minor and fixable (incorrect answer formatting, removable extra tags, or rewritable description mentions), set \texttt{"format\_valid": true} and place the minimally corrected version in \texttt{"fixed"}. If violations are severe (multiple conflicting answers, unremovable dependence on the description), set \texttt{"format\_valid": false} and \texttt{"fixed": null}.
\end{promptbox}
\caption{Stage~4 prompt used with $M_{\text{instruct}}$ to verify and (where possible) minimally repair long reasoning traces generated via thought continuation.}
\label{fig:prompt-verify-long}
\end{figure*}

\section{Training Details}
\label{appen:training_details}
\input{3_method/table_dataset_stats}
For SFT, we perform full fine-tuning of all model components, including the LLM, the audio encoder, and the projection layers. We use a learning rate of $6 \times 10^{-6}$, batch size of 256, and train the model for 5 epochs. All SFT experiments are implemented using LLaMAFactory \cite{llamafactory}.

Direct Preference Optimization (DPO) \cite{rafailov2023direct} learns the policy $\pi_\theta$ from pairwise preferences by increasing the likelihood of a preferred response $o_w$ over a dispreferred response $o_l$, while regularizing against a fixed reference policy $\pi_{\mathrm{ref}}$ via a log-ratio contrast, as shown in Eq.~\ref{eq:DPO-obj-seq}.

\begin{figure*}[t]
\begin{equation}
\mathcal{J}_{\mathrm{DPO}}(\theta)
= \mathbb{E}_{(q,o_w,o_l)\sim \mathcal{D}}
\left[
\log \sigma\!\left(
\beta \log \frac{\pi_\theta(o_w \mid q)}{\pi_{\mathrm{ref}}(o_w \mid q)}
-\beta \log \frac{\pi_\theta(o_l \mid q)}{\pi_{\mathrm{ref}}(o_l \mid q)}
\right)
\right].
\label{eq:DPO-obj-seq}
\end{equation}
\end{figure*}

We train DPO using a learning rate of $5 \times 10^{-7}$ for 3 epochs with a batch size of 256. We use the sigmoid function $\sigma(\cdot)$ with a $\beta = 0.2$. We additionally include the SFT loss with a weight of 1.0 to stabilize training following \citet{rpa}. All DPO experiments are implemented using Swift~\cite{swift}.

GRPO  \cite{shao2024deepseekmath} uses the mean reward of responses sampled from the policy model as the baseline when computing the advantage.
Specifically, a group of responses $\{o_1, o_2, \cdots, o_G\}$ is sampled from the policy model given the input question $q$, which yields corresponding rewards $\{r_1, r_2, \cdots, r_G\}$ from the reward model. The advantage is calculated over the mean:
\begin{equation}
    \hat{A}_{i, t} = \widetilde{r}_i = \frac{r_i - \mathrm{mean}(\mathbf{r})}{\mathrm{std}(\mathbf{r})}.
\end{equation}

The policy model is further regularized with a Kullback--Leibler (KL) divergence penalty, as shown in Eq.~\ref{eq:GRPO-obj}.

\begin{figure*}[t]
\begin{equation}
\begin{split}
    \mathcal{J}_{\mathrm{GRPO}}(\theta) ={}& \mathbb{E}_{q \sim P(Q),\, \{o_i\}_{i=1}^G \sim \pi_{\theta_{\mathrm{old}}}(O \mid q)}
    \Bigg[ \frac{1}{G} \sum_{i=1}^G \frac{1}{|o_i|} \sum_{t=1}^{|o_i|} \bigg( \min \Big[
    \frac{\pi_\theta(o_{i,t} \mid q, o_{i,<t})}{\pi_{\theta_{\mathrm{old}}}(o_{i,t} \mid q, o_{i,<t})} \hat{A}_{i,t}, \\
    & \mathrm{clip}\Big( \frac{\pi_\theta(o_{i,t} \mid q, o_{i,<t})}{\pi_{\theta_{\mathrm{old}}}(o_{i,t} \mid q, o_{i,<t})}, 1 - \epsilon, 1 + \epsilon \Big)  \hat{A}_{i,t}
    \Big] - \beta \mathbb{D}_{\mathrm{KL}}\!\left[\pi_{\theta} \,\|\, \pi_{\mathrm{ref}}\right] \bigg) \Bigg]
\end{split}
\label{eq:GRPO-obj}
\end{equation}
\end{figure*}

Following common practice \cite{deepseek}, we adopt a rule-based reward scheme that combines an accuracy reward for correct MCQ answers and a format reward for outputs that follow the \texttt{<think></think>} and \texttt{<answer></answer>} structure.
The accuracy reward is assigned a weight of 1.0, while the format reward is weighted by 0.05.

We use a learning rate of $1.5 \times 10^{-6}$. For each prompt, we sample $G=8$ responses from the policy model, set the batch size to 64, and train for 1,000 optimization steps. We use a sampling temperature of 1.0 and set the KL regularization coefficient $\beta$ to 0.005. All GRPO experiments are implemented using Swift~\cite{swift}.

For every training, experiments are conducted on a single node with 8 NVIDIA H100 GPUs, and all reported results are from a single run.

\section{Evaluation Details}
\label{appen:evaluation_details}
We mainly evaluate models on the widely used MMAU Test-mini~\cite{mmau}, MMAR~\cite{mmar}, and MMSU~\cite{mmsu} benchmarks, which assess general audio understanding and reasoning capabilities across diverse tasks.
MMAU evaluates information extraction and reasoning across speech, sound events, and music.
MMAR focuses on complex, multi-step reasoning tasks spanning speech, sound, music, and their combinations.
MMSU places greater emphasis on speech, evaluating both perceptual and reasoning abilities across paralinguistic and linguistic dimensions.
For all benchmarks, we follow the official evaluation protocols: string matching is used for MMAU and MMAR, and letter matching is used for MMSU, with answers extracted from the \texttt{<answer></answer>} tags. To ensure fair comparison with previously reported results, we fix the benchmark version used for each: the \texttt{v05.15.25} release for MMAU Test-mini, and for MMAR and MMSU the annotations released before February 1, 2026, prior to subsequent updates to both benchmarks.

\input{7_appendix/artifact_statement}

\section{Use of AI Assistants}
We used AI assistants (large language models) only for minor coding support and language editing. All research ideas, experiments, analyses, and claims are the authors' own, and the authors reviewed and verified all AI-assisted code and text.

%% file: 7_appendix/table_correction.tex
\begin{table}[h]
\centering
\small
\resizebox{\columnwidth}{!}{\begin{tabular}{lcc}
\hline
\textbf{Pre $\rightarrow$ Post} & \textbf{\# Responses} & \textbf{\%} \\
\hline
correct $\rightarrow$ correct                               & 668 & 59.0 \\
incorrect $\rightarrow$ incorrect                           & 309 & 27.3 \\
incorrect $\rightarrow$ correct \textit{(correction)}       & 99  & 8.7  \\
correct $\rightarrow$ incorrect \textit{(over-correction)}  & 56  & 4.9  \\
\hline
\textbf{Net correction} & \textbf{+43} & \textbf{+3.8} \\
\hline
\end{tabular}}
\caption{Pre-trigger vs.~post-trigger answer transitions on MMAR and MMAU responses from the two-stage model (LLT SFT~$\rightarrow$~DPO) that contain a Stage~4 cognitive trigger, over the 1{,}132 responses with a determinable pre-trigger answer.
The net correction is statistically significant (McNemar test, $p{<}0.001$).}
\label{tab:cognitive_correction}
\end{table}

%% file: 7_appendix/table_ablation_stage4.tex
\begin{table}[t]
\centering
\resizebox{\linewidth}{!}{%
\begin{tabular}{lccc}
\hline
\textbf{Setting} &
\makecell[c]{\textbf{MMAU} \\ \textbf{Test-mini}} &
\textbf{MMAR} &
\textbf{MMSU} \\
\hline
Thought expansion     & 75.6 & 60.5 & 63.8 \\
Thought continuation  & \textbf{75.7} & \textbf{61.1} & \textbf{64.0} \\
\hline
\end{tabular}%
}
\caption{Ablation on long-trace construction (Stage~4). The best result is in bold.}
\label{tab:ablation_stage4}
\end{table}

%% file: 7_appendix/table_nll.tex
\begin{table}[t]
\centering
\resizebox{\linewidth}{!}{%
\begin{tabular}{lcc}
\hline
\textbf{Trace (role in DPO)} & \textbf{Base} & \makecell[c]{\textbf{SFT ckpt} \\ \textbf{(DPO init)}} \\
\hline
$r^{+}$ (chosen)                                   & 2.173 & 0.715 \\
$r^{+}\!\rightarrow\! r^{++}$ (chosen)             & 2.153 & 0.831 \\
$r^{-}\!\rightarrow\! r^{++}$, synthetic (chosen)  & 2.348 & 1.428 \\
$r^{-}\!\rightarrow\! r^{++}$, model-sampled (chosen) & 1.774 & 1.282 \\
$r^{-}$, synthetic (rejected)                      & 2.510 & 1.948 \\
$r^{-}$, model-sampled (rejected)                  & 0.767 & 1.615 \\
\hline
\end{tabular}%
}
\caption{Per-token NLL of each trace type (lower = more likely under the model).}
\label{tab:nll}
\end{table}

%% file: 7_appendix/table_nll_separation.tex
\begin{table}[t]
\centering
\resizebox{\linewidth}{!}{%
\begin{tabular}{lcc}
\hline
\textbf{Preference pair} & \textbf{Synthetic} & \textbf{Model-sampled} \\
\hline
$r^{+} \succ r^{-}$                                  & 1.233 & 0.900 \\
$r^{+}\!\rightarrow\! r^{++} \succ r^{-}$            & 1.117 & 0.784 \\
$r^{-}\!\rightarrow\! r^{++} \succ r^{-}$            & 0.520 & 0.333 \\
\hline
\end{tabular}%
}
\caption{Separation between chosen and rejected responses under the DPO reference policy.}
\label{tab:nll_separation}
\end{table}

%% file: 7_appendix/table_mmau_pro.tex
\begin{table}[t]
\centering
\resizebox{\linewidth}{!}{%
\begin{tabular}{lccc}
\hline
\textbf{Model} & \textbf{Overall} & \makecell[c]{$\leq$\textbf{30s} \\ ($n{=}1720$)} & \makecell[c]{$>$\textbf{30s} \\ ($n{=}2862$)} \\
\hline
Qwen2.5-Omni-7B (reported)     & 52.2 & --   & --   \\
Qwen2.5-Omni-7B (reproduced)   & 50.1 & 42.6 & 57.1 \\
\quad + LLT (SFT $\rightarrow$ DPO) & \textbf{55.2} & \textbf{48.7} & \textbf{59.8} \\
\hline
\end{tabular}%
}
\caption{Results on MMAU-Pro.}
\label{tab:mmau_pro}
\end{table}

%% file: 7_appendix/table_medmosaic.tex
\begin{table*}[t]
\centering
\resizebox{\textwidth}{!}{%
\begin{tabular}{lccccccccc}
\hline
\textbf{Model} & \makecell[c]{\textbf{Weighted} \\ \textbf{Avg}} & \textbf{Long-Form} & \textbf{Cough} & \textbf{Heart} & \textbf{Lung} & \textbf{Speech} & \makecell[c]{\textbf{Speech} \\ \textbf{+ Sound}} & \textbf{Multi-Turn} & \textbf{Voice QA} \\
\hline
Qwen2.5-Omni-7B                     & 47.5 & \textbf{54.7} & 44.4 & 38.0 & 36.1 & \textbf{55.6} & 46.1 & \textbf{100.0} & \textbf{56.1} \\
\quad + LLT (SFT $\rightarrow$ DPO) & \textbf{51.3} & 45.3 & \textbf{51.1} & \textbf{45.8} & \textbf{44.3} & 55.3 & \textbf{52.8} & 88.9 & 47.8 \\
\hline
\end{tabular}%
}
\caption{Results on the publicly released subset of MedMosaic.}
\label{tab:medmosaic}
\end{table*}

%% file: 7_appendix/table_wow_bench.tex
\begin{table}[t]
\centering
\resizebox{\linewidth}{!}{%
\begin{tabular}{lccc}
\hline
\textbf{Model} & \textbf{Overall} & \textbf{Perception} & \textbf{Cognition} \\
\hline
Qwen2.5-Omni-7B                     & 35.9 & 38.9 & 35.5 \\
\quad + LLT (SFT $\rightarrow$ DPO) & \textbf{40.4} & \textbf{44.6} & \textbf{39.6} \\
\hline
\end{tabular}%
}
\caption{Results on WoW-Bench.}
\label{tab:wow_bench}
\end{table}

%% file: 7_appendix/table_audio_ablation.tex
\begin{table}[t]
\centering
\resizebox{\linewidth}{!}{%
\begin{tabular}{lcc}
\hline
\textbf{Model} & \makecell[c]{\textbf{MMAU} \\ \textbf{Test-mini}} & \textbf{MMAR} \\
\hline
Qwen2.5-Omni-7B (reported)   & 71.5 & 56.7 \\
Qwen2.5-Omni-7B (reproduced) & 74.6 & 57.9 \\
\quad + Gaussian noise       & 49.2 & 36.9 \\
\quad + No audio             & 48.4 & 34.6 \\
\hline
LLT (SFT $\rightarrow$ DPO)  & 78.2 & 63.7 \\
\quad + Gaussian noise       & 48.9 & 31.4 \\
\quad + No audio             & 58.7 & 39.3 \\
\hline
\end{tabular}%
}
\caption{Audio ablation on MMAU Test-mini and MMAR.}
\label{tab:audio_ablation}
\end{table}

%% file: 7_appendix/figure_interface.tex
\begin{figure*}[t]
    \centering
    \includegraphics[width=\textwidth]{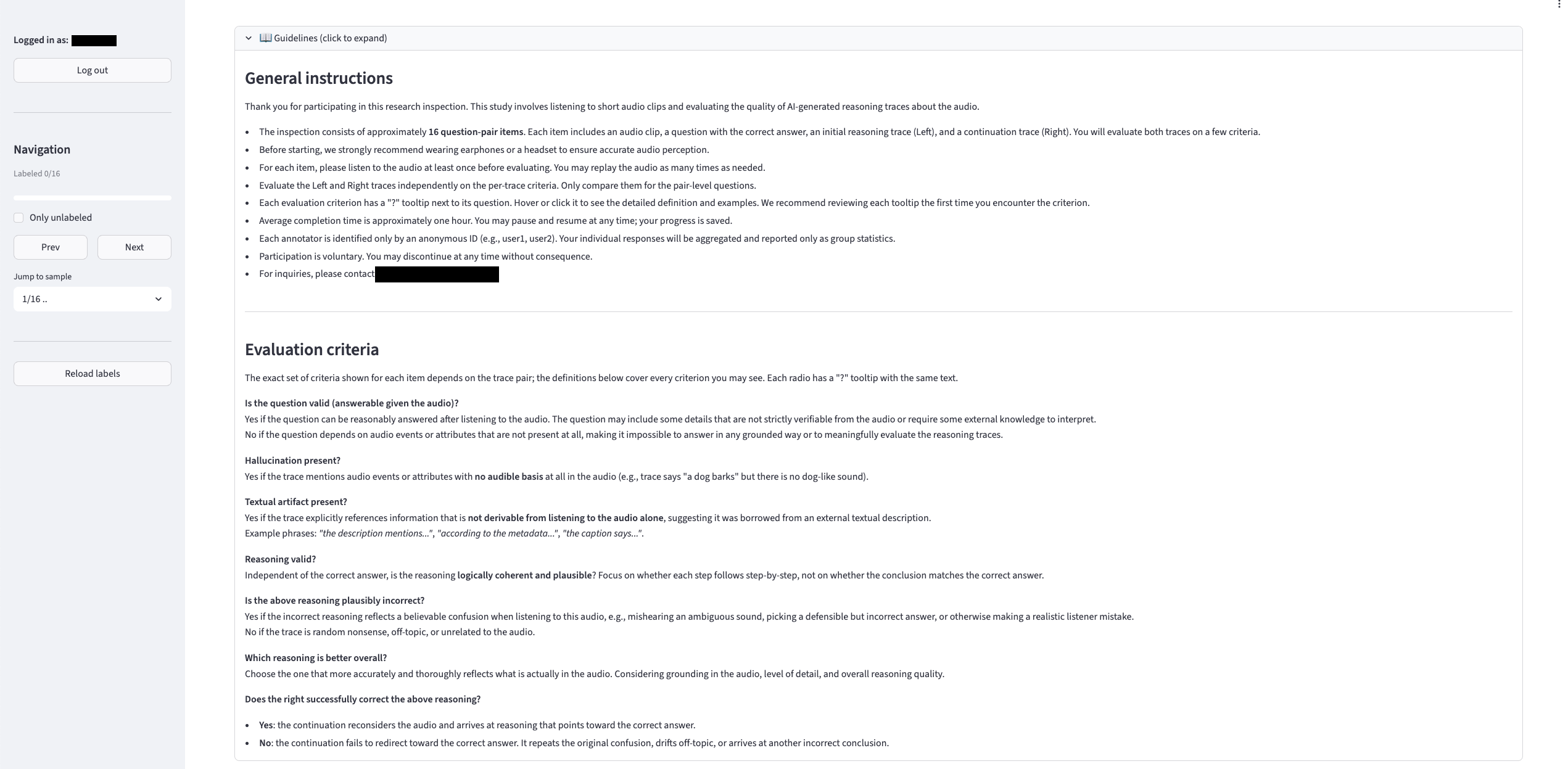}
    \caption{Full annotation guidelines shown to annotators at the start of the labeling interface.}
    \label{fig:annotation_guidelines}
\end{figure*}

\begin{figure*}[t]
    \centering
    \begin{subfigure}{\textwidth}
        \centering
        \includegraphics[width=\textwidth]{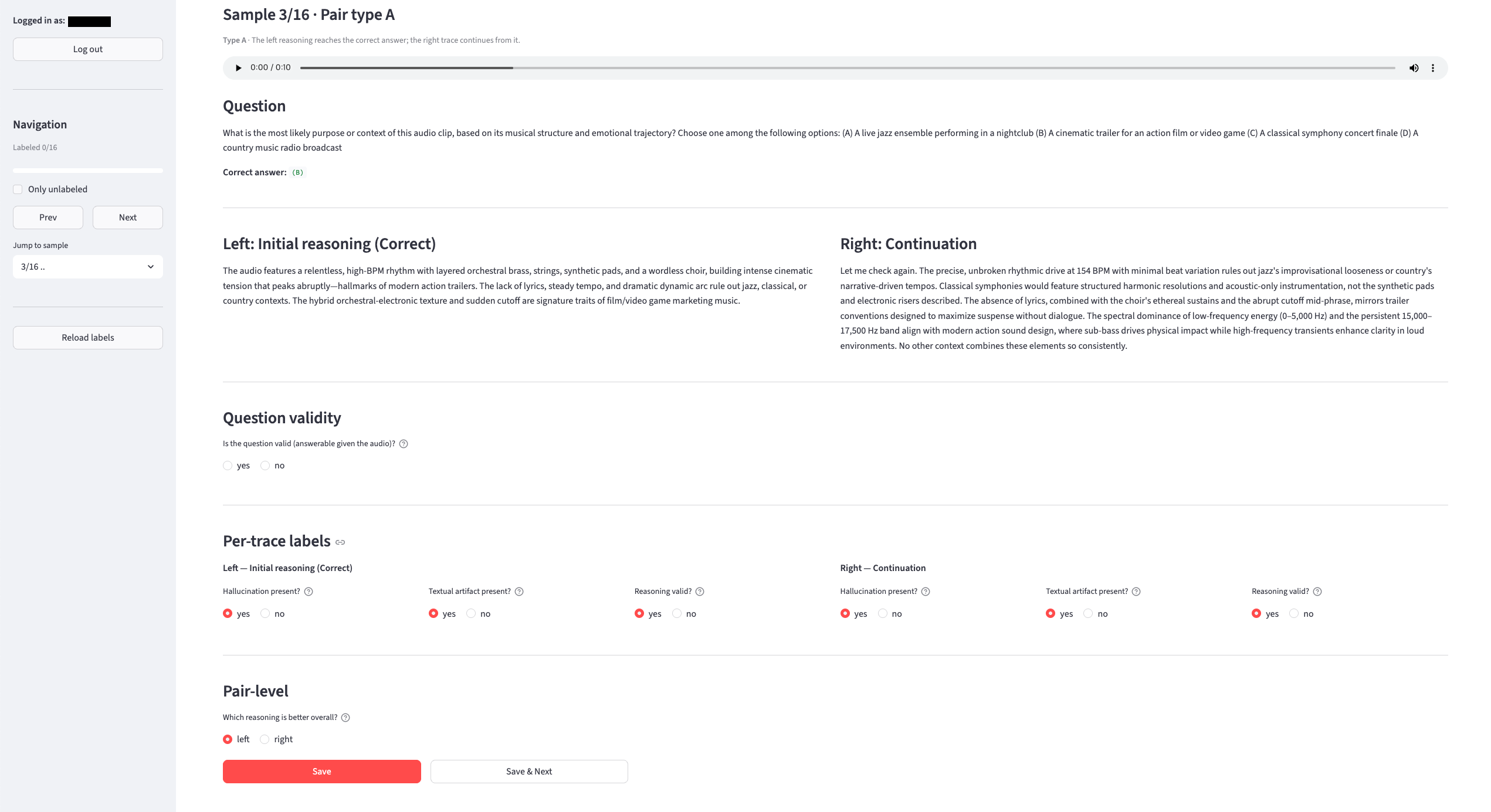}
        \caption{Type~A pair $(r^{+},\, r^{+}\!\rightarrow\! r^{++})$.}
        \label{fig:annotation_interface_a}
    \end{subfigure}
    \\[6pt]
    \begin{subfigure}{\textwidth}
        \centering
        \includegraphics[width=\textwidth]{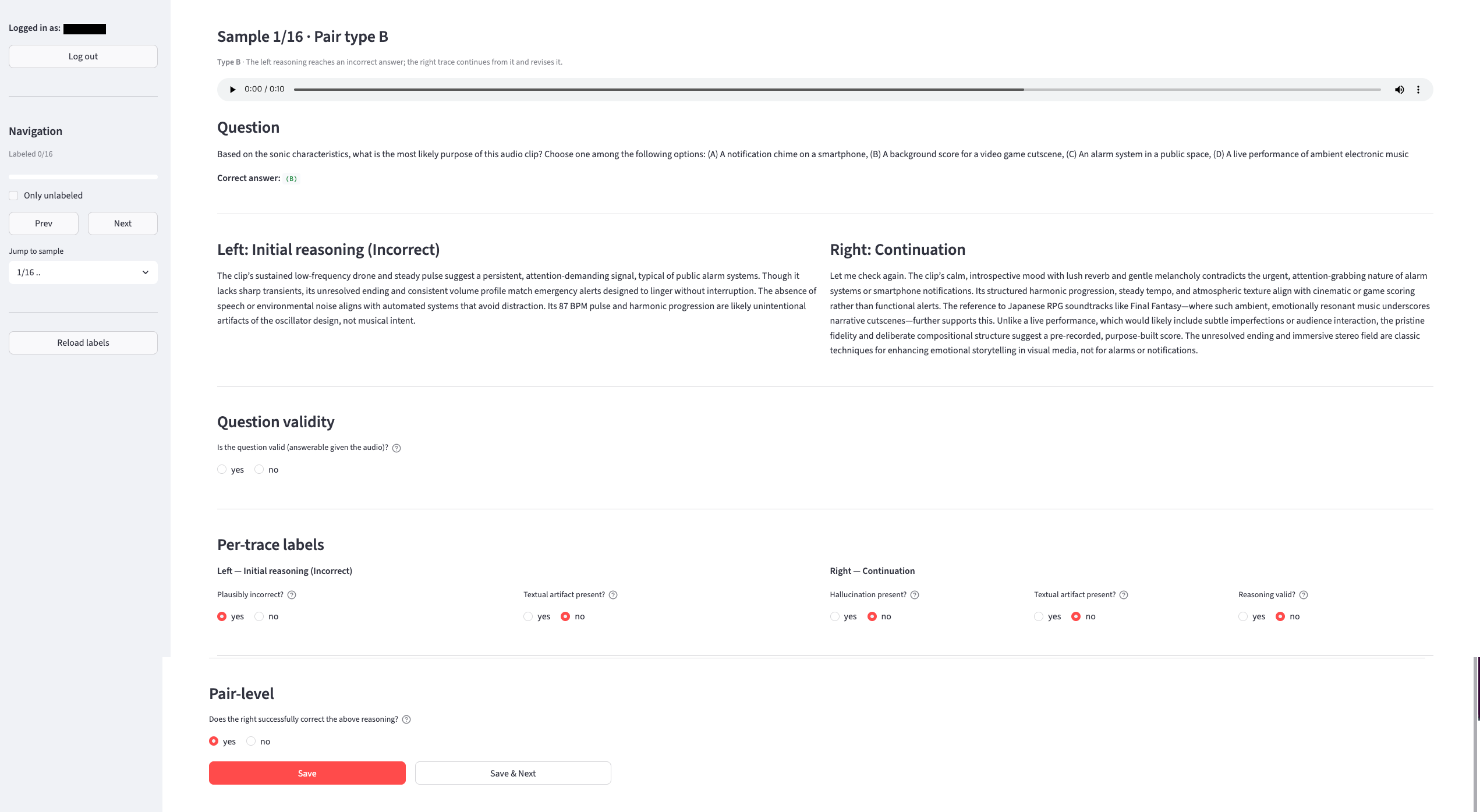}
        \caption{Type~B pair $(r^{-},\, r^{-}\!\rightarrow\! r^{++})$.}
        \label{fig:annotation_interface_b}
    \end{subfigure}
    \caption{Labeling interface presented to annotators for (a) Type~A and (b) Type~B trace pairs. Each item shows the audio clip, the question with its correct answer, the initial trace (left) and the continuation (right), and the per-trace and pair-level judgments.}
    \label{fig:annotation_interface}
\end{figure*}

%% file: 7_appendix/table_human_eval.tex
\begin{table}[t]
\centering
\small
\setlength{\tabcolsep}{4pt}
\resizebox{0.9\columnwidth}{!}{%
\begin{tabular}{lcccc}
\hline
& \multicolumn{2}{c}{\textbf{Type A}} & \multicolumn{2}{c}{\textbf{Type B}} \\
\cmidrule(lr){2-3}\cmidrule(lr){4-5}
& $r^{+}$ & $r^{++}$ & $r^{-}$ & $r^{++}$ \\
\hline
\multicolumn{5}{l}{\textit{Per-trace quality}} \\
Hallucination          & 13 & 19 & --  & 18 \\
Textual artifact       & 1  & 1  & 1   & 2  \\
Reasoning valid        & 96 & 99 & --  & 95 \\
Plausibly incorrect    & -- & -- & 77  & -- \\
\hline
\multicolumn{5}{l}{\textit{Pair-level judgment}} \\
Better reasoning       & 35 & \textbf{65} & -- & -- \\
Successful correction  & -- & -- & \multicolumn{2}{c}{\textbf{93}} \\
\hline
\end{tabular}%
}
\caption{Human inspection results (\% of annotations on the valid-question subset, $N{=}172$). Question validity is 90\% overall (Type~A 88\%, Type~B 92\%).}
\label{tab:human_eval}
\end{table}

\begin{table}[t]
\centering
\small
\resizebox{\columnwidth}{!}{\begin{tabular}{lccc}
\hline
\textbf{Item} & \textbf{\% Agr.} & \textbf{Cohen's $\kappa$} & \textbf{Krippendorff's $\alpha$} \\
\hline
Question valid         & 85  & 0.22    & 0.22 \\
Hallucination          & 79  & 0.21    & 0.22 \\
Textual artifact       & 97  & $-0.01$ & $-0.01$ \\
Reasoning valid        & 94  & $-0.03$ & $-0.03$ \\
Plausibly incorrect    & 66  & 0.07    & 0.09 \\
Better reasoning       & 50  & $-0.09$ & $-0.08$ \\
Successful correction  & 88  & $-0.06$ & $-0.05$ \\
\hline
\end{tabular}}
\caption{Inter-annotator agreement over the 96 doubly-annotated pairs.}
\label{tab:human_eval_iaa}
\end{table}

%% file: 7_appendix/figure_spectrogram.tex
\begin{figure}[t]
    \centering
    \includegraphics[width=\linewidth]{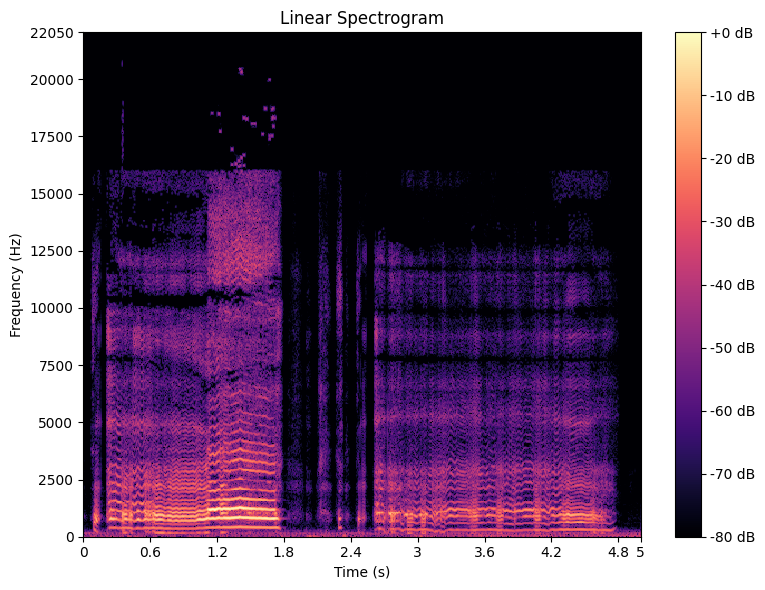}
    \caption{Spectrogram of \texttt{ESC-50/3-151081-B-20} (Baby Crying)}
    \label{fig:appen_spectrogram}
\end{figure}

%% file: 3_method/table_dataset_stats.tex
\begin{table}[t]
\centering
\small
\begin{tabular}{lr}
\hline
\textbf{Source data} &  \\
\quad Audio clips & 338k \\
\quad Original QA pairs (AF-Think) & 338k \\
\quad Hardened QA pairs (Stage~2) & 316k \\
\quad Total QA pairs & 654k \\
\hline
\textbf{Reasoning traces} & \\
\quad Short correct $r^{+}$ & 654k \\
\quad Correct-to-refine $r^{+}\!\rightarrow\! r^{++}$ & 505k \\
\quad Incorrect-to-correct $r^{-}\!\rightarrow\! r^{++}$ & 441k \\
\quad Total & 1.6M \\
\hline
\textbf{Two-stage training} & \\
\quad SFT ($r^{+}$, $r^{+}\!\rightarrow\! r^{++}$) & 1.16M \\
\quad DPO preference pairs ($20\text{k}\times3$) & 60k \\
\hline
\end{tabular}
\caption{LLT dataset statistics. Counts are approximate.}
\label{tab:dataset_stats}
\end{table}

%% file: 7_appendix/artifact_statement.tex
\section{Scientific Artifacts}
\label{appen:artifacts}
Our work builds on open-source models and publicly available datasets, and we release the LLT dataset, the trained models, and the full data-construction pipeline.
We enumerate the artifacts used and created below.
To the best of our knowledge, our use of all artifacts is consistent with their intended research use, and we release our artifacts under terms compatible with those of the source data (research use only).
License details are noted where available.

\subsection{Models}
\begin{itemize}
  \item \textbf{Qwen2.5-Omni-7B}~\cite{qwen25_omni}\footnote{\url{https://github.com/QwenLM/Qwen2.5-Omni}} --- base LALM that we fine-tune (Apache-2.0).
  \item \textbf{Qwen3-Next-80B-Instruct}\footnote{\url{https://huggingface.co/Qwen/Qwen3-Next-80B-A3B-Instruct}} / \textbf{Qwen3-Next-80B-Thinking}\footnote{\url{https://huggingface.co/Qwen/Qwen3-Next-80B-A3B-Thinking}} --- $M_{\text{instruct}}$ / $M_{\text{reason}}$ for data construction (Apache-2.0).
  \item \textbf{Qwen3-Omni-Captioner}~\cite{qwen3_omni}\footnote{\url{https://huggingface.co/Qwen/Qwen3-Omni-30B-A3B-Captioner}} --- semantic descriptions in Stage~1 (Apache-2.0).
  \item \textbf{Qwen3-VL-32B}~\cite{qwen3_vl}\footnote{\url{https://huggingface.co/Qwen/Qwen3-VL-32B-Instruct}} --- spectrogram interpretation in Stage~1 (Apache-2.0).
  \item \textbf{Qwen3-32B}~\cite{qwen3}\footnote{\url{https://huggingface.co/Qwen/Qwen3-32B}} --- text reasoner used only for the Stage-1 analysis (Apache-2.0).
  \item \textbf{Qwen2-Audio-Instruct}~\cite{qwen2_audio}\footnote{\url{https://huggingface.co/Qwen/Qwen2-Audio-7B-Instruct}} / \textbf{Qwen3-Omni-Instruct}~\cite{qwen3_omni}\footnote{\url{https://huggingface.co/Qwen/Qwen3-Omni-30B-A3B-Instruct}} --- additional backbones for the generalization study (Apache-2.0).
  \item \textbf{faster-whisper}~\cite{whisper}\footnote{\url{https://github.com/SYSTRAN/faster-whisper}} (MIT), \textbf{pyannote.audio}~\cite{pyannote}\footnote{\url{https://github.com/pyannote/pyannote-audio}} (MIT), and a \textbf{BEATs}-based sound event detector~\cite{beats, pretrained_sed}\footnote{\url{https://github.com/fschmid56/PretrainedSED}} (MIT) --- temporal information in Stage~1.
  \item \textbf{allin1}~\cite{allin1}\footnote{\url{https://github.com/mir-aidj/all-in-one}} (MIT) and \textbf{autochord}~\cite{autochord}\footnote{\url{https://github.com/cjbayron/autochord}} (MIT) --- music structure and chord recognition in Stage~1.
  \item \textbf{LLaMA-Factory}~\cite{llamafactory}\footnote{\url{https://github.com/hiyouga/LLaMA-Factory}} (Apache-2.0) for SFT and \textbf{ms-swift}~\cite{swift}\footnote{\url{https://github.com/modelscope/ms-swift}} (Apache-2.0) for DPO/GRPO; \textbf{LoRA}~\cite{lora} for the generalization study.
  \item \textbf{GPT-4.1-mini}\footnote{\url{https://platform.openai.com/docs/models/gpt-4.1-mini}} --- cognitive behavior classification and correction analysis in Section~4.2 (proprietary; OpenAI Terms of Use).
\end{itemize}

\subsection{Datasets}
\begin{itemize}
  \item \textbf{AF-Think}~\cite{af3}\footnote{\url{https://huggingface.co/datasets/nvidia/AF-Think}} --- the source audio CoT dataset we enrich (NVIDIA OneWay Noncommercial License). For reproducibility we exclude licensed subsets such as \textbf{Music4all}~\cite{music4all} and limit audio to 30 seconds. The underlying audio originates from public datasets (e.g., LibriSpeech, VGGSound, AudioSet, Clotho, DailyTalk, MELD, FMA, ESC-50).
\item \textbf{AVQA}~\cite{avqa}\footnote{\url{https://github.com/AlyssaYoung/AVQA}} --- audio question-answering data for the GRPO stage.
\item \textbf{AudioMCQ}~\cite{audio_mcq}\footnote{\url{https://huggingface.co/datasets/inclusionAI/AudioMCQ}} --- alternative source dataset for the generalization study (Apache-2.0).
  \item \textbf{MMAU}~\cite{mmau}, \textbf{MMAR}~\cite{mmar}, \textbf{MMSU}~\cite{mmsu}, \textbf{WoW-Bench}~\cite{wow_bench}, and \textbf{TREA}~\cite{trea} --- evaluation benchmarks.
\end{itemize}

\subsection{Released Artifacts}
We release (i)~the \textbf{LLT dataset} of approximately 1.6M reasoning traces (statistics in Table~\ref{tab:dataset_stats}), (ii)~the \textbf{LLT-trained models}, and (iii)~the full \textbf{data-construction pipeline}.
These are released for research use, consistent with the research-use terms of the source datasets; derivatives of research-only data are not intended for use outside research contexts.

\subsection{Data Statistics, Coverage, and Privacy}
The LLT dataset spans speech, sound events, and music; per-component statistics and training splits are reported in Table~\ref{tab:dataset_stats} and Appendix~\ref{appen:training_details}.
All audio and question--answer pairs are drawn from existing public research datasets; we do not collect new data or add personally identifying metadata.
Speech recordings in these datasets may contain individual voices, but we introduce no new identifying annotations.
The human inspection (Appendix~\ref{app:human_eval}) was conducted by annotators identified only by anonymous IDs, with all responses aggregated into group statistics.